\documentclass[twocolumn]{IEEEtran}

\usepackage{graphicx}
\usepackage{array}
\usepackage{color}
\usepackage{cite}
\usepackage{epstopdf}
\usepackage{float}
\usepackage{times,amssymb,amsmath}
\usepackage{amsfonts}
\usepackage{lineno}
\usepackage{subcaption}

\begin{document}
\title{Tracking performance of RLS algorithms \\ in WSSUS channels} 
\author{Y.~Zakharov, \IEEEmembership{Senior Member, IEEE} and L.~Shen, \IEEEmembership{Member, IEEE} \\ ~\\
\small School of Physics, Engineering and Technology, University of York, UK   
}
\markboth{Draft}%
{\quad}
\maketitle

\begin{abstract}
Adaptive algorithms are widely used for estimation of linear time-varying systems, such as communication channels. Their tracking performance depends on the level of noise, characteristics of time variations, and the algorithm parameters. Optimizing these parameters and predicting the algorithm performance is an important task. In this paper, we present an approach for analysing the tracking performance of recursive least squares (RLS) adaptive algorithms in channels with time variations described as wide-sense stationary uncorrelated scattering (WSSUS) random processes characterised by a power spectral density (PSD). We focus on exponential RLS and sliding-window RLS (SRLS) algorithms, for which general formulas for the mean square deviation (MSD) as a measure of the tracking performance are obtained in terms of spectral moments of the PSD. As examples, they are specified for random processes with uniform, Jakes' and autoregressive PSDs. These results are further generalized to RLS algorithms with delays, e.g. processing delays or delays introduced in non-causal adaptive RLS algorithms, and to the SRLS-L algorithm with approximation of channel time variations using Legendre polynomials. Numerical examples show good match between the analytical MSD and simulation results.  
\end{abstract}

\IEEEpeerreviewmaketitle

\section{Introduction} 

Recursive least squares (RLS) adaptive algorithms (filters) are widely used for identification of linear systems, such as communications channels~\cite{lin1995optimal}. They have been applied to channel estimation in multi-antenna communication systems~\cite{karami2007tracking},  underwater acoustic communication systems~\cite{yeo2000improved, zhang2018soft, zakharov2016sliding}, full-duplex systems~\cite{shen2022bem,shen2019digital}, etc. The estimation performance depends on the level of noise, characteristics of time variations and algorithm parameters. Optimizing the parameters and prediction of the algorithm performance is an important task, which has been well addressed in the literature. 

In most works, the tracking performance of the exponential-window RLS (ERLS) algorithm is studied for channel variations described by the first-order Markov process~\cite{eleftheriou1986tracking,bershad1990performance, guo1992tracking, eweda1994comparison, karami2007tracking, silva2008improving, claser2021tracking}.  
In~\cite{eleftheriou1986tracking}, the excess lag mean squared error (MSE) is derived for the ERLS adaptive filter. In~\cite{eweda1994comparison}, closed-form expressions for the steady-state excess MSE and mean squared deviation (MSD) for the ERLS algorithm are obtained. 

A more practical channel model for mobile communication applications is the wide-sense stationary uncorrelated scattering (WSSUS) channel with Jakes' power spectral density (PSD) of the channel tap variations~\cite{karami2007tracking, lin1995optimal}. The work in~\cite{karami2007tracking} considers approximation of Jakes' model by an autoregressive (AR) model to simplify the analysis of the ERLS tracking performance. In~\cite{filimon1993lms}, the MSD (called the mean square identification error in~\cite{filimon1993lms} and~\cite{lin1995optimal}) is obtained for the ERLS algorithm as an integral of the channel scattering function, which is difficult to analyse and use in practice. In~\cite{lin1995optimal}, a unified expression is obtained for calculating the MSD for ERLS and sliding-window RLS (SRLS) algorithms. This is given in a form of an integral of PSD of channel tap variations and specified into a simple expression for Jakes' PSD. However, this expression does not take into account an important error component, the least squares (LS) modelling component, and therefore, as will be shown in our paper, it is not applicable to adaptive algorithms with improved tracking performance, such as non-causal RLS algorithms with delays or to non-causal RLS algorithms with approximation of time variations by basis functions~\cite{niedzwiecki2019generalized, shen2020adaptive, shen2021performance, shen2022bem}, e.g., Legendre polynomials in the SRLS-L algorithm~\cite{shen2020adaptive, shen2021performance}. In~\cite{shen2021performance}, we derived a general MSD expression for the SRLS and SRLS-L algorithms, which is however difficult to analyse and use in practice.  

A delay ($t_0$) in an adaptive filter is the time shift between the time instant of the last sample used in the RLS estimation and the time to which this estimate is assigned. The delay $t_0$ can be zero, positive or negative. The case $t_0 = 0$ corresponds to classical RLS algorithms.  The case $t_0 < 0$ corresponds to a processing delay when implementing the adaptive filter; in other words, the channel estimate is used later than the last received signal sample involved in computation of this estimate; as a result, the tracking performance degrades. The case $t_0 > 0$ corresponds to non-causal RLS adaptive filters, when the impulse response estimate is assigned to a previous time instant; this can be acceptable in communication systems, but it results in a latency in data transmission. In the case $t_0 > 0$, the tracking performance can be significantly improved compared to the case $t_0 = 0$~\cite{niedzwiecki2019generalized, shen2020adaptive, shen2021performance, shen2022bem}. 

In this paper, we analyse the tracking performance of SRLS and ERLS adaptive filters with and without delay in WSSUS channels and compare the theoretical results with corresponding simulation results. The contributions of the paper are as follows. 
\begin{itemize}
\item An approach is proposed for analysing the tracking performance of RLS algorithms in channels with time variations described as WSSUS random processes. 
\item Based on this approach, general simple formulas for prediction of tracking performance of classical RLS adaptive filters in WSSUS channels are derived for arbitrary PSD, which are then specified for uniform, Jakes' and AR PSDs. 
\item The results are generalized to RLS algorithms with delays, such as the delayed SRLS (dSRLS) and delayed ERLS (dERLS) algorithms, and to the SRLS-L algorithm. 
\item Numerical simulation is conducted, which shows good match between analytical and simulation results.
\end{itemize}

The paper is organized as follows. Section~\ref{Sec:Channel-Model-Methodology} describes the channel model and methodology of the analysis. In Sections~\ref{Sec:Tracking-SRLS},~\ref{Sec:Tracking-SRLS-L} and ~\ref{Sec:Tracking-ERLS}, we analyse the tracking performance of the SRLS, SRLS-L and ERLS 	adaptive filters, respectively. In Section~\ref{Sec:Numerical-results} we compare the tracking performance obtained analytically and in simulation. Section~\ref{Sec:Conclusions} concludes the paper.

\section{Channel model and methodology} \label{Sec:Channel-Model-Methodology}

In this section, the linear channel model is introduced in subsection~\ref{Subsec:Channel-Model}. Characteristics of the channel time variations are described in subsection~\ref{Subsec:Channel-Variations}. Methods to generate channel realizations in the simulation are described in subsection~\ref{Subsec:Channel-Simulation}. The analysis methodology adopted in the paper is described in subsection~\ref{Subsec:Methodology}.

\subsection{Channel model} \label{Subsec:Channel-Model}

We consider the channel model described by the discrete-time relationship: 
\begin{eqnarray} \label{Eq:Linear_channel}
x(i) =  \mathbf{h}^T(i) \mathbf{s}(i) + n(i) , 
\end{eqnarray}
where $\mathbf{s}(i) = [s(i), s(i-1), \ldots, s(i-L+1)]^T$ is an $L~\times~1$ vector, $s(i)$ is the signal at the input of the channel with a time-varying impulse response $\mathbf{h}(i)$ to be estimated, and $n(i)$ is a noise signal. The input signal $s(i)$ and noise $n(i)$ are sequences of independent zero-mean Gaussian random numbers with variances $\sigma_s^2$ and $\sigma_n^2$, respectively. We assume that elements (impulse response taps) $h_\ell(i)$, $\ell = 0, \ldots, L-1$, of the $L \times 1$ vector $\mathbf{h}(i)$ are samples of realizations of continuous-time random processes $h_\ell(t)$: $h_\ell(i) = h_\ell(i T_s)$, where $T_s$ is a sampling interval. More specifically, $\{h_\ell(t)\}$ are independent zero-mean stationary Gaussian random processes. To simplify the notation, we   write $h_\ell(i)$ to denote a discrete-time tap, while $h_\ell(t)$ denotes the corresponding continuous-time tap. To simplify the presentation, we assume that all variables in~(\ref{Eq:Linear_channel}) are real-valued.       

\subsection{Characteristics of channel time variations} \label{Subsec:Channel-Variations}

The process $h_\ell(t)$ is defined by the normalized PSD $G_h(\omega)$:  
\begin{eqnarray} \label{Eq:Gh_norm}
\frac{1}{2 \pi} \int_{-\infty}^{\infty} G_h(\omega) d \omega = 1.  
\end{eqnarray}
The impulse response $\mathbf{h}(t)~=~[h_0(t),\ldots,h_\ell(t),\ldots,h_{L-1}(t)]^T$ is also characterised by the power delay profile (PDP) $\{ P_\ell \}$, $\ell = 0, \ldots , L-1$, which is an $L$-length vector of variances of corresponding channel taps, $P_\ell = E\{h_\ell^2(t) \}$; two examples will be used for the numerical simulation: the uniform PDP ($P_\ell = 1$) and triangle PDP ($P_\ell = 1 - \ell / L$). 

For the analysis, we will be using the spectral moments 
\begin{eqnarray} \label{Eq:Gh_spectral_moments}
\lambda_{2p} = \frac{1}{2 \pi} \int_{-\infty}^{\infty} \omega^{2p} G_h(\omega) d \omega, \ \ p = 0, 1, 2 \ldots . 
\end{eqnarray}
Note that $P_\ell \lambda_{2p}$ is the variance of the $p$th derivative $h_\ell^{(p)}(t)$ of $h_\ell(t)$, i.e., $P_\ell \lambda_{2p} =  E\{ [ h_\ell^{(p)}(t)]^2 \}$ ~\cite{cramer2013stationary,leadbetter2012extremes}. 
We will arrive at expressions for modelling and approximation errors as functions of the spectral moments using the Taylor series for representation of the time variant tap in the vicinity of $t = 0$:   
\begin{eqnarray} \label{Eq:h_Taylor}
h_\ell(t) =  \sum_{p = 0}^{\infty} \frac{h_\ell^{(p)}(0)}{p!} t^p . 
\end{eqnarray}
Note that the use of the spectral moments for analysing the accuracy of approximation of random processes is a well known approach~\cite{cramer2013stationary, leadbetter2012extremes, zakharov2004polynomial}.

Three specific PSDs will be considered as examples: uniform, Jakes' and AR PSD. The uniform PSD is given by  
\begin{eqnarray} \label{Eq:Uniform_Gh}
G_h(\omega) = \left\{ \begin{array}{ll}
(2 f_\text{max})^{-1},  &|\omega| < 2 \pi f_\text{max} , \\
0 , &\text{otherwise} ,
\end{array}
\right.
\end{eqnarray}
where $f_\text{max}$ is the maximum frequency of the random process, and the corresponding spectral moments are 
\begin{eqnarray} \label{Eq:Uniform_Gh_spectral_moments}
\lambda_{2p} = \frac{(2\pi f_\text{max})^{2p}}{2p+1}, \ \lambda_{2} = \frac{4(\pi f_\text{max})^{2}}{3}, \ \lambda_{4} = \frac{16(\pi f_\text{max})^{4}}{5}.
\end{eqnarray}

For Jakes' PSD, often used to describe time-variations of mobile wireless communication channels~\cite{feng2008statistical}, we have  
\begin{eqnarray} \label{Eq:Clarkes_Gh}
G_h(\omega) = \left\{ \begin{array}{ll}
\left[ \pi f_\text{max} \sqrt{1 - \left( \frac{\omega}{2 \pi f_\text{max}} \right)^2} \right]^{-1} ,  &|\omega| < 2 \pi f_\text{max} , \\
0 , &\text{otherwise} ,
\end{array}
\right.
\end{eqnarray}
and the corresponding spectral moments are 
\begin{eqnarray} \label{Eq:Clarks_Gh_spectral_moments}
\lambda_{2p} &=& \frac{ (2p)! (\pi f_\text{max})^{2p}}{ (p!)^2 }, \nonumber \\ 
\lambda_{2} &=& 2(\pi f_\text{max})^{2}, \ \lambda_{4} = 6(\pi f_\text{max})^{4}. 
\end{eqnarray}

The AR PSD is given by
\begin{eqnarray} \label{Eq:AR_Gh}
G_h(\omega) = \frac{2 \mu}{\mu^2 + \omega^2} ,
\end{eqnarray}
where the parameter $\mu$ characterises the frequency bandwidth of the random process. Note that, for this process, the spectral moments in~(\ref{Eq:Gh_spectral_moments}) do not exist (for $p>0$, the integral is infinity). However, a discrete-time version of the process (see subsection~\ref{Subsec:Channel-Simulation}) is often used for analysing the tracking performance of adaptive filters and therefore we will also consider this case in our analysis.

\subsection{Generating random channels in the numerical simulation} \label{Subsec:Channel-Simulation}

In the numerical simulation, realizations of a random process describing the time-varying channel taps will be generated as 
\begin{eqnarray}
h_\ell(t) = \sqrt{P_\ell} A \cos(\omega t + \varphi) ,
\label{Eq:Ch-Sim-Cos}
\end{eqnarray}                                              
where $A$, $\omega$ and $\varphi$ are independent random variables.  The probability density functions of $A$ and $\varphi$ are given by 
\begin{eqnarray}
p_A(A) &=& A \exp \left\{ -\frac{A^2}{2} \right\}, \ \ A \geq 0 , 
\label{Eq:Ch-Sim-Densities-A}  \\
p_\varphi(\varphi) &=& \frac{1}{2\pi}, \ \  -\pi <\varphi \leq \pi ,  
\label{Eq:Ch-Sim-Densities-phi}
\end{eqnarray}                                              
respectively, i.e., $A$ has the Rayleigh distribution and $\varphi$ has the uniform distribution. In this case, $h_\ell(t)$ in~(\ref{Eq:Ch-Sim-Cos}) is a zero-mean Gaussian random process with variance $E\{h_\ell^2(t)\} = P_\ell$. 

For achieving a specific PSD $G_h(\omega)$, the random variable $\omega$ should have the probability density function $(2 \pi)^{-1}G_h(\omega)$~\cite{leadbetter2012extremes}. For the uniform PSD in~(\ref{Eq:Uniform_Gh}), $\omega$ is uniformly distributed in the interval $[-2\pi f_\text{max}, 2\pi f_\text{max}]$. For Jakes' PSD in~(\ref{Eq:Clarkes_Gh}), the random variable $\omega$ is computed as $\omega = 2 \pi f_\text{max} \sin(u)$, where $u$ is a random variable uniformly distributed within $[-\pi/2, \pi/2]$. 
   
There are other methods to generate realizations of random processes with predefined PSDs, e.g., the frequency-domain methods~\cite{percival1993simulating, brillinger1974fourier} or some methods to generate Jakes' fading channels~\cite{zheng2002improved, xiao2002second, zheng2003simulation}. These methods however are approximate, i.e., statistical characteristics of the generated random processes are not exactly the same as predefined, e.g., in~(\ref{Eq:Uniform_Gh}) or~(\ref{Eq:Clarkes_Gh}). This can make validation of the analytically derived formulas inaccurate, whereas the method described above guarantees exact statistical characteristics as predefined. We emphasize however that the channel realizations in~(\ref{Eq:Ch-Sim-Cos}) are only used for the numerical simulation, they are not used in the analytical analysis. 

The AR process with the PSD~(\ref{Eq:AR_Gh}) corresponds to generating the channel taps according to 
\begin{eqnarray} 
h_\ell(t) = \mu \int_{0}^{\infty} e^{-\mu \tau} z_\ell(t - \tau) d\tau,   \label{Eq:PSD_mu_channel_taps} 
\end{eqnarray}
where $z_\ell(t)$ is a white Gaussian process and $\mu > 0$. This is a continuous-time equivalent of generating the channel taps in the discrete time as the first-order Markov process: 
\begin{eqnarray} 
\mathbf{h}(i) = \beta \mathbf{h}(i-1) + \mathbf{z}(i),   \label{Eq:PSD_mu_channel_taps_discrete} 
\end{eqnarray}
where $\mathbf{z}(i)$ is a vector of samples of $\{z_\ell(t)\}$ at $t = iT_s$ and $\beta = e^{-\mu T_s}$. This channel model is used in many works (e.g., see~\cite{haykin2002adaptive, eweda1994comparison, eleftheriou1986tracking}) for analysis of the tracking performance of the classical ERLS algorithm. Therefore, we will also consider it in our analysis.

\subsection{Methodology} \label{Subsec:Methodology}

We consider an adaptive filter as an estimator of the channel impulse response $\mathbf{h}(t)$. Since the channel is described as a stationary random process, in our analysis, we can consider any time instant, e.g. $t = 0$; we also assume that the adaptive filter was initialized at $t = -\infty$, i.e. the transient period of the adaptive filter has finished by the time $t = 0$. 

The performance of the adaptive filter will be measured in terms of the normalized MSD
\begin{eqnarray} \label{Eq:NMSD}
\text{NMSD} = \frac{E\{ || \mathbf{h}(0) - \hat{\mathbf{h}}(0) ||^2 \}}{E\{ || \mathbf{h}(0)||^2 \}}  = \frac{1}{P} \sum_{\ell = 0}^{L-1} E\{ e_\ell^2(0) \} ,    
\end{eqnarray}
where $P = \sum_{\ell = 0}^{L-1} P_\ell$, $\hat{\mathbf{h}}(0)$ is an estimate of $\mathbf{h}(0)$, $e_\ell(t) = h_\ell(t) - \hat{h}_\ell(t)$ is the estimation error of the $\ell$th tap and $\hat{h}_\ell(t)$ is the $\ell$th tap estimate. 

Assuming that the noise $n(i)$ is independent of the channel response $\mathbf{h}(i)$ and the input signal $s(i)$, the NMSD can be expressed as a sum of two components~(e.g., see~\cite{haykin2002adaptive}) 
\begin{eqnarray} \label{Eq:NMSD_two_components}
\text{NMSD} = \epsilon_\text{am} + \epsilon_n ,    
\end{eqnarray}
where $\epsilon_\text{am}$ is an approximation and modelling component, and $\epsilon_n$ is a noise component. The component $\epsilon_n$ depends on the noise and does not depend on the channel response. This component is given by~\cite{kay1993statistical, sayed2003fundamentals} 
\begin{eqnarray} 
\epsilon_n & \approx &  \frac{L \sigma_n^2}{M \sigma_s^2 P} = \frac{L}{M \cdot \text{SNR}} ,    \label{Eq:NMSD_epsilon_n_SRLS} \\
\epsilon_n & \approx &  \frac{(1 - \lambda) L \sigma_n^2}{(1 + \lambda) \sigma_s^2 P} =  \frac{(1 - \lambda)}{(1 + \lambda)} \cdot \frac{L}{\text{SNR}} ,    \label{Eq:NMSD_epsilon_n_ERLS}
\end{eqnarray}
for the SRLS and ERLS algorithms, respectively, where $M$ is the observation window length of the SRLS algorithm, $\lambda$ is the forgetting factor of the ERLS algorithm, and $\text{SNR} = \frac{\sigma_s^2 P}{\sigma_n^2}$ is the signal-to-noise ratio.   

The component $\epsilon_\text{am}$ depends on characteristics of the channel variations and properties of the adaptive algorithm; it does not depend on the noise. When analysing the tracking performance, we focus on finding this component, assuming that the noise $n(i)$ is absent. 
We show below that $\epsilon_\text{am}$ contains an LS modelling component $\epsilon_\text{m}$ and an approximation component $\epsilon_\text{a}$: $\epsilon_\text{am} = \epsilon_\text{m} + \epsilon_\text{a}$. 

\begin{figure}
\begin{center}  
\includegraphics[width=0.3\textwidth]{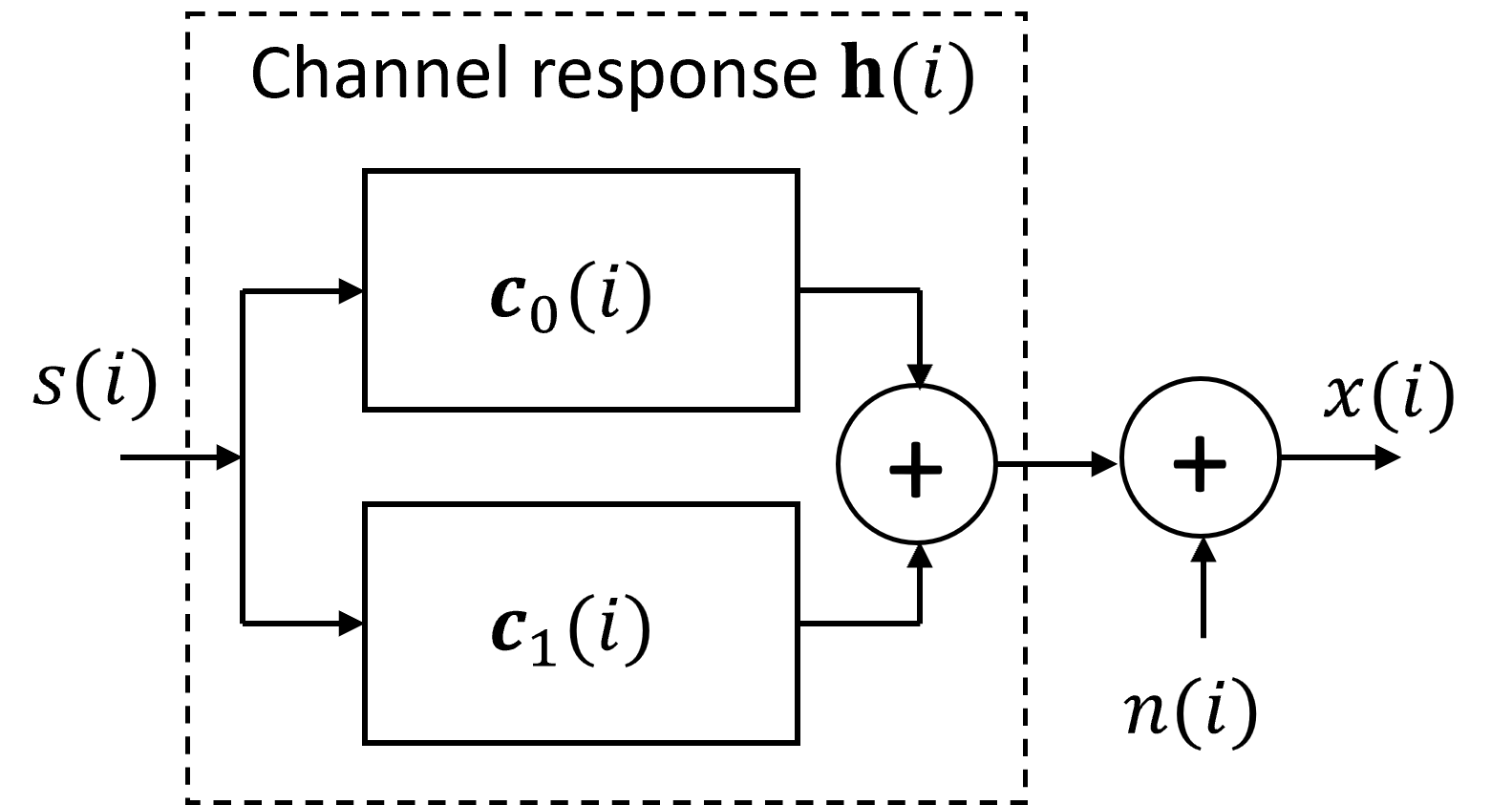}
\caption{\label{Fig:channel_model_1} Channel model. } 
\end{center}
\end{figure}
\begin{figure}
\begin{center}  
\includegraphics[width=0.3\textwidth]{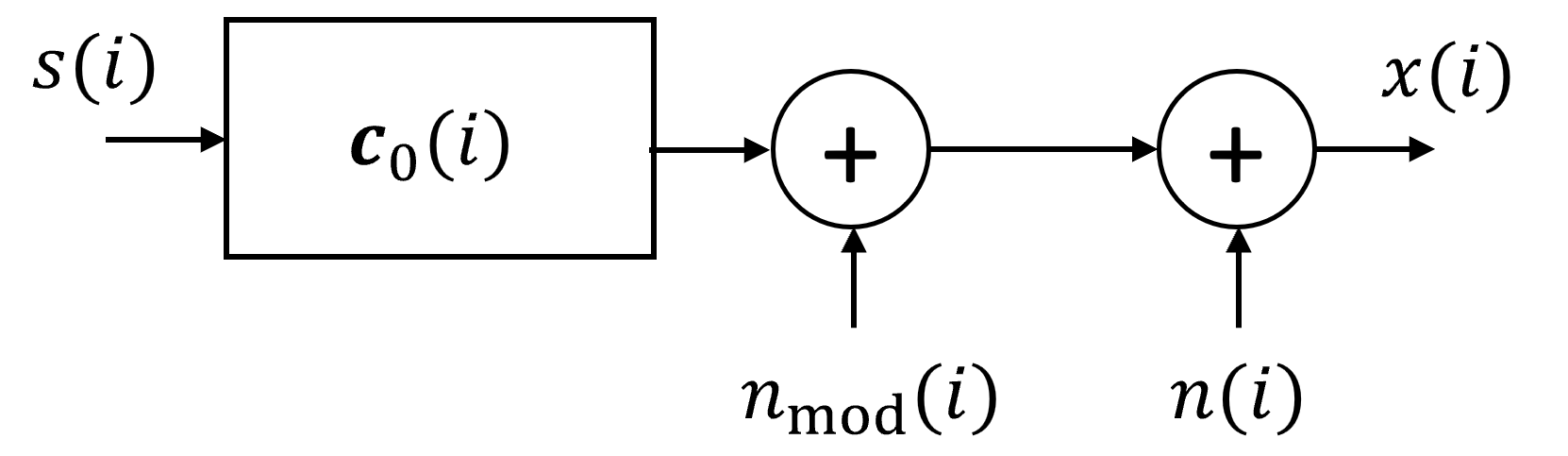}
\caption{\label{Fig:channel_model_2} Equivalent channel model. } 
\end{center}
\end{figure}
Fig.~\ref{Fig:channel_model_1} illustrates the channel model described by the time-variant channel response $\mathbf{h}(i)$. Here, we assume that, at every time instant $i$, the channel is described by two vectors, $\mathbf{c}_0(i)$ and $\mathbf{c}_1(i)$; $\mathbf{c}_0(i)$ is a vector of parameters estimated by an LS algorithm, implemented as an adaptive filter, and $\mathbf{c}_1(i)$ is a vector of parameters (nuisance parameters) not covered by the LS signal model. If $\mathbf{c}_1(i)$ is a non-zero vector, there is a modelling error, which can be transformed into a modelling noise $n_\text{mod}(i)$ as shown in Fig.~\ref{Fig:channel_model_2}. The modelling noise is a result of filtering the input signal $s(i)$ by a filter described by the parameters~$\mathbf{c}_1(i)$. 

The modelling noise in an observation time window $w(t)$, over which the LS estimation is performed, can be represented as 
\begin{eqnarray} \label{Eq:Modelling_noise}
n_\text{mod}(t) =  \sum_{\ell = 0}^{L-1} \tilde{h}_\ell(t) s(t - \ell T_s),   
\end{eqnarray}
where $\tilde{h}_\ell(t)$ is the channel impulse response component, which is not covered by the LS signal model. We 
will assume that the time variant tap $h_\ell(t)$ in the observation interval can be represented as the linear combination 
\begin{eqnarray} \label{Eq:h_two_basis_functions}
h_\ell(t) =  c_{0, \ell} L_0(t) + c_{1, \ell} L_1(t), 
\end{eqnarray}
where $L_0(t)$ and $L_1(t)$ are two orthonormal basis functions with respect to the window $w(t)$ and $c_{0,\ell}$ and $c_{1,\ell}$ are corresponding expansion coefficients: 
\begin{eqnarray} \label{Eq:Expansion_coefficients} 
c_{\ell,q} = \int_{-\infty}^{\infty} w(t) h_\ell(t) L_q(t) dt , \ \ \ q = 0, 1.  
\end{eqnarray}
The window $w(t)$ is rectangular for SRLS algorithms and exponential for ERLS algorithms.          
The parameters $\{c_{0,\ell}\}$ are covered by the LS signal model~\cite{kay1993statistical}, whereas parameters $\{c_{1,\ell}\}$ are not. In this case, $\tilde{h}_\ell(t) = c_{1,\ell} L_1(t)$. Then~(\ref{Eq:Modelling_noise}) can be rewritten as
\begin{eqnarray} \label{Eq:Modelling_noise-2}
n_\text{mod}(t) =  L_1(t) \sum_{\ell = 0}^{L-1} c_{1, \ell} s(t - \ell T_s) .   
\end{eqnarray}
It can be seen that in general the modelling noise $n_\text{mod}(t)$ is non-stationary due to the factor $L_1(t)$. To simplify our derivation, we replace the non-stationary noise by an equivalent stationary noise with variance equal to the average variance of $n_\text{mod}(t)$ over the observation interval. 

The variance of $n_\text{mod}(t)$ is given by 
\begin{eqnarray} \label{Eq:Mod-noise-variance}
E\{ n_\text{mod}^2(t)\} =  \sigma_s^2 L_1^2 (t) \sum_{\ell = 0}^{L-1} E\{c_{1, \ell}^2 \} .    
\end{eqnarray}
Here we used the independence of the channel components $h_\ell(t)$ and the channel input $s(t)$. 
The average variance is found as  
\begin{eqnarray} \label{Eq:Mod-noise-average-variance}
\sigma^2_\text{mod} &=&  \frac{1}{|| w^2 ||} \int_{-\infty}^{\infty} w^2(t)  E\{ n_\text{mod}^2(t)\} dt \nonumber \\ 
&=& k_w \sigma_s^2 \sum_{\ell = 0}^{L-1} E\{c_{1, \ell}^2 \}  ,    
\end{eqnarray}
where $k_w = || w^2 ||^{-1} \int_{-\infty}^{\infty} w^2(t) L_1^2(t)dt$ and $||w^2|| = \int_{-\infty}^{\infty} w^2(t) dt$.   
Then the modelling NMSD component can be estimated using formulas~(\ref{Eq:NMSD_epsilon_n_SRLS}) and~(\ref{Eq:NMSD_epsilon_n_ERLS}) and replacing $\sigma_n^2$ with $\sigma_\text{mod}^2$ from~(\ref{Eq:Mod-noise-average-variance}) for SRLS and ERLS algorithms, respectively. 

Let us consider the classical ERLS or SRLS adaptive filter. If the channel is time-invariant, i.e. $\mathbf{h}(i) = \mathbf{h}$, and the channel length $L$ is the same as the length of the vector $\mathbf{c}_0(i)$, so that $\mathbf{c}_0(i) = \mathbf{h}$, we obtain that $\mathbf{c}_1(i)$ is a zero vector and there is no modelling error: $\epsilon_\text{m} = 0$. Therefore, in the absence of the noise $n(i)$ and with a proper input signal $s(i)$, the ERLS or SRLS algorithm can solve the LS problem in finding the vector $\mathbf{c}_0(i) = \mathbf{h}$ without errors. If the channel is time-variant, the vector $\mathbf{c}_1(i)$ is non-zero, thus resulting in a non-zero modelling component $\epsilon_\text{m} > 0$. However, now the vector $\mathbf{c}_0(i)$ estimated by the LS algorithm is not the same as $\mathbf{h}(i)$: $\mathbf{c}_0(i) \neq \mathbf{h}(i)$; this inequality results in a non-zero approximation component~$\epsilon_\text{a}~>~0$. 

If an LS estimator estimates both the vectors $\mathbf{c}_0(i)$ and $\mathbf{c}_1(i)$, then the modelling error is zero, but the approximation error may still be present. This will be further discussed when considering the SRLS-L adaptive filter in Section~\ref{Sec:Tracking-SRLS-L}.

\section{Tracking performance of SRLS adaptive filters} \label{Sec:Tracking-SRLS}

In this section, we first consider the classical SRLS adaptive algorithm presented in Table~\ref{Tab:SRLS}. 
The estimate of the impulse response $\hat{\mathbf{h}}(i)$ at every time instant $i$ is found by solving the system of equations $\mathbf{R}(i) \hat{\mathbf{h}}(i) = \mathbf{b}(i)$, where $\mathbf{R}(i) = \mathbf{S}^T(i) \mathbf{S}(i)$ is the $L \times L$ autocorrelation matrix of the channel input and $\mathbf{b}(i) = \mathbf{S}^T(i) \mathbf{x}(i)$ is the $L \times 1$ vector of cross-correlation between the channel input and output,  $\mathbf{S}(i) = [\mathbf{s}(i), \ldots , \mathbf{s}(i-M+1)]^T$ is an $M \times L$ matrix of the input signal, $\mathbf{x}(i) = [x(i), x(i-1), \ldots, x(i-M+1)]^T$ is an $M~\times~1$ observed signal vector, and $M$ is the length of the sliding window. 
\begin{table}[t]
\centering
\caption{Classical SRLS algorithm \label{Tab:SRLS}}
\begin{tabular}{ll}
\hline
\textbf{Step} & \textbf{Equation} \\ \hline
     & for every $i > -\infty$, repeat: \\
  1  & \ \ \ \ \ $\mathbf{b}(i) = \mathbf{S}^T(i) \mathbf{x}(i)$ \\
  2  & \ \ \ \ \ $\mathbf{R}(i) = \mathbf{S}^T(i) \mathbf{S}(i) $ \\
  3  & \ \ \ \ \ Solve the system: $\mathbf{R}(i) \hat{\mathbf{h}}(i) = \mathbf{b}(i)$ \\
  4  & \ \ \ \ \ Output of the algorithm: $\hat{\mathbf{h}}(i)$ \\
      \hline
\end{tabular}
\end{table}

We then consider a more general case of the delayed SRLS (dSRLS) adaptive algorithm (see Table~\ref{Tab:dSRLS}) with a delay $i_0 = t_0/T_s$, assuming that the ratio $t_0 / T_s$ is an integer number (negative, zero or positive). The only difference between the dSRLS and SRLS algorithms is the delay $i_0$ in the processed signals $x(i)$ and $s(i)$. In the case $i_0 = 0$, the dSRLS algorithm is transformed into the SRLS algorithm.  
\begin{table}[t]
\centering
\caption{Delayed SRLS (dSRLS) algorithm \label{Tab:dSRLS}}
\begin{tabular}{ll}
\hline
\textbf{Step} & \textbf{Equation} \\ \hline
     & for every $i > -\infty$, repeat: \\
  1  & \ \ \ \ \ $\mathbf{b}(i) = \mathbf{S}^T(i+i_0) \mathbf{x}(i+i_0)$ \\
  2  & \ \ \ \ \ $\mathbf{R}(i) = \mathbf{S}^T(i+i_0) \mathbf{S}(i+i_0) $ \\
  3  & \ \ \ \ \ Solve the system: $\mathbf{R}(i) \hat{\mathbf{h}}(i) = \mathbf{b}(i)$ \\
  4  & \ \ \ \ \ Output of the algorithm: $\hat{\mathbf{h}}(i)$ \\
      \hline
\end{tabular}
\end{table}

\subsection{Modelling error of SRLS algorithms}
  
For SRLS algorithms, the observation window is $w(t) = T^{-1}$ if $t \in [-T/2, T/2]$ and zero otherwise. We choose $L_0(t) = 1$ and $L_1(t) = 2\sqrt{3} ( t / T)$, i.e. they are, respectively, zero and first order orthonormal Legendre polynomials~\cite{gradshteyn2014table} defined over the interval $[-T/2,T/2]$, while $c_{0,\ell}$ and $c_{1,\ell}$ are given by
\begin{eqnarray} \label{Eq:h_Legendre_expansion coefficients}
c_{q,\ell} = \frac{1}{T} \int_{-T/2}^{T/2} h_\ell(t) L_q(t) dt  , \ \ q = 0,1 .
\end{eqnarray}
Note that an SRLS algorithm solves the LS problem defined by the signal model $h_\ell(t) =  c_{0,\ell} L_0(t)$ and thus it estimates parameters $c_{0,\ell}$.  
If $c_{1,\ell} = 0$ for $\ell = 0,\ldots,L-1$, i.e., $\mathbf{h}(t)$ is time-invariant within $[-T/2,T/2]$, there is no LS modelling error and the output of the SRLS algorithm, in the absence of the additive noise $n(i)$, is a vector with elements $c_{0,\ell}$, so that $\hat{h}_\ell(0) = c_{0,\ell} L_0(0) = h_\ell(0) $.

If $c_{1,\ell} \neq 0$, there will be a modelling error $\tilde{h}_\ell = c_{1,\ell} L_1(t)$. Since $h_\ell(t)$ is a random process, then $c_{1,\ell}$ is a random variable. Using the Taylor series in~(\ref{Eq:h_Taylor}), from~(\ref{Eq:Expansion_coefficients}) we obtain
\begin{eqnarray}  \label{Eq:h_Legendre_first_c_1}
c_{1,\ell} = h_\ell^{(1)}(0) \frac{1}{T} \int_{-T/2}^{T/2} t \cdot L_1(t) dt = \frac{T}{\sqrt{12}} \ h_\ell^{(1)}(0). 
\end{eqnarray}
Finally, from~(\ref{Eq:Mod-noise-average-variance}), using $k_w = 1$ and the relationship $E\{[h_\ell^{(1)}(0)]^2 \} = P_\ell \lambda_2$, we arrive at the average variance of the modelling error,  
\begin{eqnarray}
\sigma^2_\text{mod} = P \sigma_s^2 \frac{\lambda_2 T^2}{12} .   
\end{eqnarray}
It is seen that $\sigma^2_\text{mod}$ increases with the speed of the channel variations manifested by the spectral moment $\lambda_2$. As $\lambda_2 \rightarrow 0$, the extra noise due to the LS modelling error vanishes.     

Using~(\ref{Eq:NMSD_epsilon_n_SRLS}) (and replacing $\sigma_n^2$ with $\sigma^2_\text{mod}$), the contribution of the modelling noise into the NMSD can be computed as
\begin{eqnarray} \label{Eq:NMSD_SRLS_modelling_component}
\epsilon_\text{m}^\text{(SRLS)} & \approx &  \frac{\lambda_2 T^2}{12} \left( \frac{L}{M} \right). 
\end{eqnarray}
This formula and its variants for the uniform and Jakes' PSDs are shown in Table~\ref{Table:SRLS_NMSD_components}, where we denote $\gamma = f_\text{max} T$. We can also write $\gamma = f_m M$, where $f_m = f_\text{max} T_s$ is the bandwidth $f_\text{max}$ of the channel variations normalized to the sampling frequency $1/T_s$. 

The analysis of the modelling component of the NMSD has been done assuming that the observation interval is $[-T/2,T/2]$. However, since the random process describing the channel variations is stationary, this result is the same for other (time-shifted) intervals of length $T$. This result is equally applicable to the classical SRLS and dSRLS algorithms, which estimate the vector $\mathbf{c}_0$ with elements $\{c_{0,\ell}\}$ in~(\ref{Eq:h_Legendre_expansion coefficients}) or their time-shifted version. However, $\mathbf{c}_0$ is only an approximation of the channel impulse response and thus there is an approximation error that we need to analyse.  
 
\subsection{Approximation error of SRLS algorithms}
 
We will now derive the variance of the approximation error in terms of the spectral moments $\lambda_{2p}$ assuming that the additive and modelling noises are not present. We assume that the estimate $\hat{h}_\ell(0)$ of a single tap is given by  
\begin{eqnarray} \label{Eq:SRLS_h_hat_integral}
\hat{h}_\ell(0) =  \frac{1}{T} \int_{t_0-T}^{t_0} h_\ell(t) dt, 
\end{eqnarray}
where $t_0$ is a delay. In the classical SRLS algorithm, the delay $t_0 = 0$. However, we will be considering here a more general case of arbitrary delay $t_0$. 
In the dSRLS algorithm considered in~\cite{shen2020adaptive,shen2021performance,shen2022bem}, $t_0 = T/2$, and we show below that this is an optimal delay. 

Using the Taylor series~(\ref{Eq:h_Taylor}), the estimate in~(\ref{Eq:SRLS_h_hat_integral}) is represented as 
\begin{eqnarray} \label{Eq:SRLS_h_hat}
\hat{h}_\ell(0) =  \frac{1}{T} \sum_{p = 0}^{\infty} \frac{h_\ell^{(p)}(0)}{(p+1)!} \left[t_0^{p+1} - (t_0 - T)^{p+1} \right].  
\end{eqnarray}
In~(\ref{Eq:SRLS_h_hat}), the term $p = 0$ is equal to $h_\ell(0)$, the true value of the channel impulse response tap. Let $\eta = t_0 / T$ be the delay in terms of the sliding window length $T$, then the estimation error $e_{\text{dSRLS},\ell} = \hat{h}_\ell(0) - h_\ell(0)$ can be represented as  
\begin{eqnarray} \label{Eq:SRLS_h_hat_error}
e_{\text{dSRLS},\ell} =  \sum_{p = 1}^{\infty} \frac{h_\ell^{(p)}(0) T^p}{(p+1)!} \varphi_{p+1}(\eta),  
\end{eqnarray}
where $\varphi_p(\eta) = \eta^{p} - (\eta - 1)^{p}$. 

\subsubsection{Approximation NMSD component for the classical SRLS algorithm}

For the classical SRLS algorithm, $\eta = 0$ and, since $\varphi_p(0) = (- 1)^{p}$, we arrive at the error
\begin{eqnarray} \label{Eq:SRLS_h_hat_eta0}
e_{\text{SRLS},\ell} &=&  \sum_{p = 1}^{\infty} \frac{h_\ell^{(p)}(0) T^p}{(p+1)!} (-1)^{p+1}   \\
     &\approx& \frac{h_\ell^{(1)}(0) T}{2} - \frac{h_\ell^{(2)}(0) T^2}{6} +  \cdots \nonumber
\end{eqnarray}
Therefore, we obtain the variance of a single tap estimate as:  
\begin{eqnarray} \label{Eq:SRLS_MSD_single_tap}
E \{ e^2_{\text{SRLS},\ell} \} \approx P_\ell \left( \frac{\lambda_2 T^2}{4} + \frac{\lambda_4 T^4}{36} \right).
\end{eqnarray}
To obtain these two terms, we take into account that a random process and its derivative are uncorrelated~\cite{leadbetter2012extremes}. 
Finally, from~(\ref{Eq:NMSD}), we arrive at the approximation NMSD component for the classical SRLS algorithm: 
\begin{eqnarray} \label{Eq:SRLS_Approximation_NMSD}
\epsilon_\text{a}^\text{(SRLS)} \approx  \frac{\lambda_2 T^2}{4} + \frac{\lambda_4 T^4}{36} .
\end{eqnarray}
This formula is applicable to a random process describing the channel variations by an arbitrary PSD with finite spectral moments $\lambda_2$ and $\lambda_4$. It can be further specified for a random process with a specific PSD $G_h(\omega)$; the corresponding formulas for the uniform and Jakes' PSDs are presented in Table~\ref{Table:SRLS_NMSD_components}. Typically, $\gamma = f_\text{max} T \ll 1$, i.e., the window length is much smaller than the correlation interval (approximately $f_\text{max}^{-1}$) of the channel variations. In this case, the first term in~(\ref{Eq:SRLS_Approximation_NMSD}) can provide an accurate prediction of the SRLS approximation NMSD component.  Note that the first term in~(\ref{Eq:SRLS_Approximation_NMSD}) can also be obtained from an integral MSD expression in~\cite{lin1995optimal}. Moreover, in~\cite{lin1995optimal}, the first term of $\epsilon_\text{a}^\text{(SRLS)}$ for Jakes' PSD has been obtained in the same form as shown in Table~\ref{Table:SRLS_NMSD_components}.     

It can also be seen in Table~\ref{Table:SRLS_NMSD_components} that, for a fixed sliding window length and maximum frequency $f_\text{max}$ of channel variations, the approximation NMSD component is higher for Jakes' PSD than for the uniform PSD. This is explained by the fact that, for Jakes' process, the time variations on average are faster due to its high PSD values concentrating closely to $f_\text{max}$ as opposed to the uniform PSD. 

From~(\ref{Eq:SRLS_Approximation_NMSD}), it can also be seen that the NMSD approximation component does not depend on the PDP of the channel.

\subsubsection{Approximation NMSD component for the dSRLS algorithm with the optimal delay $\eta = 1/2$}
For the dSRLS algorithm, we first find an optimal value of $\eta$, for which the term $p = 1$ in~(\ref{Eq:SRLS_h_hat_error}) vanishes. This optimal value is found by solving $\varphi_2(\eta) = 0$ and the solution is $\eta = 1/2$, i.e., the optimal delay $t_0 = T/2$ is a half of the sliding-window length. In this case, taking into account that $\varphi_3(1/2) = 1/4$, from~(\ref{Eq:SRLS_h_hat_error}) we obtain 
\begin{eqnarray} \label{Eq:dSRLS_e_eta_optimal}
e_{\text{dSRLS},\ell} \approx \frac{h_\ell^{(2)}(0) T^2}{24} . 
\end{eqnarray}
Then the error variance for a single tap is given by
\begin{eqnarray} \label{Eq:dSRLS_MSD_single_tap_optimal}
E \{ e^2_{\text{dSRLS},\ell} \} \approx P_\ell \left( \frac{\lambda_4 T^4}{576} \right)
\end{eqnarray}
and, from~(\ref{Eq:NMSD}), the approximation NMSD component for the dSRLS algorithm with the optimal delay $\eta = 1/2$ is 
\begin{eqnarray} \label{Eq:dSRLS_MSD_optimal}
\epsilon_\text{a}^\text{(dSRLS)} \approx  \frac{\lambda_4 T^4}{576} . 
\end{eqnarray}
Table~\ref{Table:SRLS_NMSD_components} specifies this formula for uniform and Jakes' PSDs.

\subsubsection{Approximation NMSD component for the dSRLS algorithm with an arbitrary delay $\eta$}
For other values of $\eta$, from~(\ref{Eq:SRLS_h_hat_error}) we obtain
\begin{eqnarray} \label{Eq:dSRLS_e_eta}
e_{\text{dSRLS},\ell} \approx \frac{h_\ell^{(1)}(0) T}{2} \varphi_2(\eta) + \frac{h_\ell^{(2)}(0) T^2}{3} \varphi_3(\eta)  . 
\end{eqnarray}
Then, by using derivations similar to those for the classical SRLS and optimal dSRLS algorithms, we arrive at the NMSD approximation component for the dSRLS algorithm with arbitrary delay:   
\begin{eqnarray} \label{Eq:dSRLS_MSD_arbitrary_delay}
\epsilon_\text{a}^\text{(dSRLS)} \approx \frac{\lambda_2 T^2}{4}\varphi_2^2(\eta) + \frac{\lambda_4 T^4}{36} \varphi_3^2(\eta) .
\end{eqnarray}

Table~\ref{Table:SRLS_NMSD_components} summarizes results of this section by presenting the modelling and approximation NMSD components for the SRLS algorithms.
\begin{table*}[]
\caption{NMSD modelling and approximation components of SRLS, dSRLS and SRLS-L algorithms}
\label{Table:SRLS_NMSD_components}
\centering
\begin{tabular}{|c|ccc|c|}
\hline
NMSD component         
 & \multicolumn{1}{c|}{\begin{tabular}[c]{@{}c@{}}SRLS\\ ($\eta = 0$)\end{tabular}} & \multicolumn{1}{c|}{\begin{tabular}[c]{@{}c@{}}dSRLS\\ (arbitrary $\eta$)\end{tabular}}                         & \begin{tabular}[c]{@{}c@{}} Optimal dSRLS\\ ($\eta = 0.5$)\end{tabular} &  \begin{tabular}[c]{@{}c@{}} SRLS-L \\ ($\eta = 0.5$)\end{tabular}          \\ \hline
\begin{tabular}[c]{@{}c@{}}Approximation\\ (general form)\end{tabular}  & \multicolumn{1}{c|}{$\frac{\lambda_2 T^2}{4} + \frac{\lambda_4 T^4}{36}$}        & \multicolumn{1}{c|}{$\frac{\lambda_2 T^2}{4}\varphi_2^2(\eta) + \frac{\lambda_4 T^4}{36}\varphi_3^2(\eta)$}     & $\frac{\lambda_4T^4}{576}$                                             & $\frac{\lambda_4T^4}{576}$   \\ \hline
\begin{tabular}[c]{@{}c@{}}Approximation\\ (Uniform PSD)\end{tabular}   & \multicolumn{1}{c|}{$\frac{\pi^2 \gamma^2}{3} + \frac{4 \pi^4 \gamma^4}{45}$}    & \multicolumn{1}{c|}{$\frac{\pi^2 \gamma^2 }{3}\varphi_2^2(\eta) + \frac{4\pi^4 \gamma^4}{45}\varphi_3^2(\eta)$} & $\frac{\pi^4 \gamma^4}{180}$                                           & $\frac{\pi^4 \gamma^4}{180}$ \\ \hline
\begin{tabular}[c]{@{}c@{}}Approximation \\ (Jakes' PSD)\end{tabular} & \multicolumn{1}{c|}{$\frac{\pi^2 \gamma^2}{2} + \frac{ \pi^4 \gamma^4}{6}$}     & \multicolumn{1}{c|}{$\frac{\pi^2 \gamma^2 }{2}\varphi_2^2(\eta) + \frac{\pi^4 \gamma^4}{6}\varphi_3^2(\eta)$}   & $\frac{\pi^4 \gamma^4}{96}$                                            & $\frac{\pi^4 \gamma^4}{96}$  \\ \hline
\begin{tabular}[c]{@{}c@{}}Modelling \\ (general form)\end{tabular}     & \multicolumn{3}{c|}{$\frac{\lambda_2 T^2}{12} \cdot \frac{L}{M}  $}                                                                                                                                                                                                                            & $\frac{\lambda_4 T^4}{720} \cdot \frac{L}{M} $                           \\ \hline
\begin{tabular}[c]{@{}c@{}}Modelling\\ (Uniform PSD)\end{tabular}       & \multicolumn{3}{c|}{$\frac{\pi^2 \gamma^2}{9} \cdot \frac{L}{M} $}                                                                                                                                                                                                                             & $\frac{\pi^4 \gamma^4}{225} \cdot \frac{L}{M}   $                            \\ \hline
\begin{tabular}[c]{@{}c@{}}Modelling\\ (Jakes' PSD)\end{tabular}      & \multicolumn{3}{c|}{$\frac{\pi^2 \gamma^2}{6} \cdot \frac{L}{M} $}                                                                                                                                                                                                                             & $\frac{\pi^4 \gamma^4}{120} \cdot \frac{L}{M}   $                          \\ \hline
\end{tabular}

\end{table*}

\section{Tracking performance of the SRLS-L adaptive filter} \label{Sec:Tracking-SRLS-L}

The analysis above shows that if the channel impulse response is constant over the observation interval (the time window of the adaptive filter), the LS signal model used for deriving the RLS algorithms is precise and the NMSD component $\epsilon_\text{am}$ is zero. However, this case is of no practical interest since the channel would then be time-invariant. In time-varying channels, the LS signal model is not accurate (the RLS parameters cannot accurately describe the channel variations within the observation interval). The accuracy can be improved by introducing additional parameters that make the LS signal model more accurate. 

The time-variation of the channel tap $h_\ell(t)$ on the observation interval can be represented using two Legendre polynomials $\{ L_q(t)\}_{q = 0, 1} $ (instead of one Legendre polynomial $L_0(t)$ used in the classical SRLS and dSRLS algorithms), i.e. 
\begin{eqnarray} \label{Eq:dSRLS_L_single_tap}
\hat{h}_\ell(t) = c_{0,\ell} L_0(t) + c_{1,\ell} L_1(t) , \ \ \ \ t \in [-T/2, T/2], 
\end{eqnarray}
where $c_{0,\ell}$ and $c_{1,\ell}$ are expansion coefficients, $L_0(t)$ is a constant over $[-T/2, T/2]$ and $L_1(t)$ is a linear function over $[-T/2, T/2]$ with $L_1(0) = 0$. Note that we are only interested in the estimate $\hat{h}_\ell(0)$ at time $t = 0$. Since $L_1(0) = 0$, we have $\hat{h}_\ell(t) = c_{0,\ell} L_0(t)$, i.e., we do not need $c_{1,\ell}$ to recover the estimate. However, when solving the LS problem for the model in~(\ref{Eq:dSRLS_L_single_tap}), we have to estimate both $c_{0,\ell}$ and $c_{1,\ell}$ jointly and only after these estimates are found, we can discard $c_{1,\ell}$. In the classical SRLS and dSRLS algorithms, the LS model is time-invariant: $\hat{h}_\ell(t) = c_{0,\ell} L_0(t)$ and the expansion coefficient $c_{0,\ell}$ found by solving the time-invariant LS problem differs from that found by solving the LS time-variant problem with the LS signal model in~(\ref{Eq:dSRLS_L_single_tap}). The more accurate signal model (in the absence of an additive noise) should result in a more accurate channel estimate. The representation~(\ref{Eq:dSRLS_L_single_tap}) is used in the non-causal basis-expansion model (BEM) adaptive filter based on the Legendre polynomials~\cite{shen2022bem,shen2021performance}.

\begin{table}[t]
\centering
\caption{SRLS-L algorithm\label{Tab:SRLS-L}}
\begin{tabular}{ll}
\hline
\textbf{Step} & \textbf{Equation} \\ \hline
     & Initialization: $\hat{\mathbf{h}}(i) = \mathbf{0}$ for $i<0$    \\
     & for $i > -\infty$, repeat: \\
  1  & $\mathbf{b}_{p}(i) = \mathbf{S}^H(i+i_0) \mathbf{\Phi}_p \mathbf{x}(i+i_0)$, $p=0,1$ \\
  2  & $\mathcal{R}_{p,q}(i) = \sum_{k=0}^{M-1} L_p(-T/2+kT_s)  \mathbf{R}_{i+i_0} L_q(-T/2+kT_s)$  \\
  3  & Generate matrix $\mathcal{R}(i)$ and vector $\mathbf{b}(i)$   \\
  4  & Solve system of equations $\mathcal{R}(i) \mathbf{c}(i) = \mathbf{b}(i)$ $\Rightarrow$ $[\hat{\mathbf{c}}_0(i); \hat{\mathbf{c}}_{1}(i)]$ \\
  5  & Output of the algorithm: $\hat{\mathbf{h}}(i) = \hat{\mathbf{c}}_0(i) $  \\
      \hline
\end{tabular}
\end{table}
The SRLS-L algorithm is presented in Table~\ref{Tab:SRLS-L}, where 
$\mathbf{\Phi}_p = \mathrm{diag} \{ L_p(-T/2), L_p(-T/2+T_s), \ldots , L_p(T/2) \}$ is an $M \times M$ diagonal matrix, $\mathbf{S}(i) = [\mathbf{s}(i), \ldots, \mathbf{s}(i-M+1)]^T$ is the $M\times L$ observation matrix, and $\mathbf{R}_i = \mathbf{s}(i)\mathbf{s}^T(i)$. The matrix $\mathcal{R}(i)$ is a block matrix with blocks $\mathcal{R}_{p,q}(i)$, $p,q = 0, 1$, 
$\mathbf{b}(i) = [\mathbf{b}^T_0(i); \mathbf{b}^T_{1}(i)]^T$, and $\mathbf{c}(i) = [\mathbf{c}_0(i); \mathbf{c}_{1}(i)]$. 


For the SRLS-L algorithm, the modelling NMSD component can be found using the same approach as for the SRLS algorithm. Using the Taylor series (\ref{Eq:h_Taylor}), it is straightforward to obtain that now (with $L_0(t)$ and $L_1(t)$ precisely recovering terms $p = 0$ and $p = 1$ in the Taylor series) the channel impulse response component $\tilde{h}_\ell(t)$ not covered by the LS signal model can be represented as $\tilde{h}_\ell(t) \approx c_{2,\ell} L_2(t)$, where $L_2(t) = (1/2)\sqrt{5/T}[3(2t/T)^2 - 1]$ is the second-order Legendre polynomial~\cite{gradshteyn2014table} and 
\begin{eqnarray}  \label{Eq:h_Legendre_first_c_2}
c_{2,\ell} =\frac{ h_\ell^{(2)}(0) }{T} \int_{-T/2}^{T/2} \left( \frac{t^2}{2} \right) L_2(t) dt = \sqrt{\frac{5}{T}} \frac{T^2}{60} h_\ell^{(2)}(0). 
\end{eqnarray}
Then the modelling noise~(\ref{Eq:Modelling_noise}) can be represented as
\begin{eqnarray} \label{Eq:Modelling_noise-3}
n_\text{mod}(t) =  L_2(t) \sum_{\ell = 0}^{L-1} c_{2, \ell} s(t - \ell T_s),   
\end{eqnarray}
and the average variance of the modelling noise is given by  
\begin{eqnarray} \label{Eq:Mod-noise-average-variance-L2}
\sigma^2_\text{mod} = \sigma_s^2 \sum_{\ell = 0}^{L-1} E\{c_{2, \ell}^2 \}  = \sigma_s^2 P \frac{\lambda_4 T^4}{720}.    
\end{eqnarray}
From results in~\cite{shen2021performance}, it follows that for the SRLS-L algorithm with two basis functions, $L_0(t)$ and $L_1(t)$, the noise NMSD component can still be computed using the formula~(\ref{Eq:NMSD_epsilon_n_SRLS}). Then, the modelling NMSD component for the SRLS-L algorithm is
\begin{eqnarray} \label{Eq:NMSD_SRLS-L_modelling_component}
\epsilon_\text{m}^\text{(SRLS-L)} & \approx &  \frac{\lambda_4 T^4}{720} \left( \frac{L}{M} \right). 
\end{eqnarray}


In the SRLS-L algorithm, the channel estimate is given by $\hat{h}_\ell(0) = c_{0,\ell} L_0(0)$, the same as for the dSRLS algorithm with optimal delay. Therefore, the NMSD approximation component at the optimal delay $t_0 = T/2$ (or $\eta = 0.5$) is given by 
\begin{eqnarray} \label{Eq:dSRLS_MSD_optimal_SRLS_L}
\epsilon_\text{a}^\text{(SRLS-L)} = \epsilon_\text{a}^\text{(dSRLS)} \approx  \frac{\lambda_4 T^4}{576} . 
\end{eqnarray}
Note that for other delays the expression~(\ref{Eq:dSRLS_MSD_arbitrary_delay}) can be used. 

The expressions (\ref{Eq:NMSD_SRLS-L_modelling_component}) and (\ref{Eq:dSRLS_MSD_optimal_SRLS_L}) specified for uniform and Jakes' PSDs are shown in Table~\ref{Table:SRLS_NMSD_components}.

\section{Tracking performance of ERLS adaptive filters} \label{Sec:Tracking-ERLS}

We first consider the performance of the classical ERLS algorithm (with zero delay). In the ERLS algorithm, as shown in Table~\ref{Tab:ERLS}, the estimate of the impulse response $\hat{\mathbf{h}}(i)$ at every time instant $i$ is found by solving the system of equations $\mathbf{R}(i) \hat{\mathbf{h}}(i) = \mathbf{b}(i)$, where $\mathbf{R}(i)$ is the $L \times L$ autocorrelation matrix of the channel input and $\mathbf{b}(i)$ is the $L \times 1$ vector of cross-correlation between the channel input and output; $\varepsilon > 0$ is a regularization parameter and $0 < \lambda < 1$ is the forgetting factor~\cite{sayed2003fundamentals}.  
\begin{table}[t]
\centering
\caption{ERLS algorithm \label{Tab:ERLS}}
\begin{tabular}{ll}
\hline
\textbf{Step} & \textbf{Equation} \\ \hline
     & Initialization: $\mathbf{b}(-\infty) = 0$, $\mathbf{R}(-\infty) = \varepsilon \mathbf{I}_L$      \\
     & for every $i > -\infty$, repeat: \\
  1  & \ \ \ \ \ $\mathbf{b}(i) = \lambda \mathbf{b}(i-1) + x(i) \mathbf{s}(i)$ \\
  2  & \ \ \ \ \ $\mathbf{R}(i) = \lambda \mathbf{R}(i) + \mathbf{s}(i)\mathbf{s}^T(i) $ \\
  3  & \ \ \ \ \ Solve system of equations $\mathbf{R}(i) \hat{\mathbf{h}}(i) = \mathbf{b}(i)$ \\
  4  & \ \ \ \ \ Output of the algorithm: $\hat{\mathbf{h}}(i)$ \\
      \hline
\end{tabular}
\end{table}

We will then consider a more general case of the ERLS adaptive filter (see Table~\ref{Tab:dERLS}) with a delay $i_0 = t_0/T_s$, assuming that the ratio $t_0 / T_s$ is an integer number (negative, zero or positive). The only difference of the dERLS and ERLS algorithms is the delay $i_0$ in the processed signals $x(i)$ and~$s(i)$.  
\begin{table}[t]
\centering
\caption{Delayed ERLS (dERLS) algorithm \label{Tab:dERLS}}
\begin{tabular}{ll}
\hline
\textbf{Step} & \textbf{Equation} \\ \hline
     & Initialization: $\mathbf{b}(-\infty) = 0$, $\mathbf{R}(-\infty) = \varepsilon \mathbf{I}_L$      \\
     & for every $i > -\infty$, repeat: \\
  1  & \ \ \ \ \ $\mathbf{b}(i) = \lambda \mathbf{b}(i-1) + x(i+i_0) \mathbf{s}(i+i_0)$ \\
  2  & \ \ \ \ \ $\mathbf{R}(i) = \lambda \mathbf{R}(i) + \mathbf{s}(i+i_0)\mathbf{s}^T(i+i_0) $ \\
  3  & \ \ \ \ \ Solve system of equations $\mathbf{R}(i) \hat{\mathbf{h}}(i) = \mathbf{b}(i)$ \\
  4  & \ \ \ \ \ Output of the algorithm: $\hat{\mathbf{h}}(i)$ \\
      \hline
\end{tabular}
\end{table}

\subsection{Modelling error of ERLS algorithms \label{Subsec:Mod-Error-ERLS}}

For ERLS algorithms, the observation window is $w(t) = \alpha e^{\alpha t}$, where $t \in [-\infty, 0)$, $\alpha = \ln(\lambda)/T_s$ and the basis functions $L_0(t) = 1$ and $L_1(t) = 1 + \alpha t$ are orthonormal Laguerre polynomials with respect to the window function $w(t)$~\cite{gradshteyn2014table}. Corresponding expansion coefficients are given by  
\begin{eqnarray} \label{Eq:h_Laguerre_expansion_coefficients}
c_{q,\ell} = \int_{-\infty}^{0} w(t) h_\ell(t) L_q(t) dt , \ \ \ q = 0, 1 .  
\end{eqnarray}
To compute the modelling NMSD component, we use~(\ref{Eq:Mod-noise-average-variance}). From~(\ref{Eq:h_Laguerre_expansion_coefficients}) and the relationship
\begin{eqnarray} \label{Eq:Corr_vs_PSD} 
E\{ h_\ell (t)  h_\ell (\tau) \} = \frac{P_\ell}{2\pi} \int_{-\infty}^{\infty} G_h(\omega) e^{j\omega(t-\tau)} d\omega ,   
\end{eqnarray}
we arrive at 
\begin{eqnarray}   \label{Eq:mod_error_L1} 
E\{ c_{1,\ell}^2 \} = \frac{P_\ell}{2\pi} \int_{-\infty}^{\infty} G_h(\omega) |\mathcal{L}_1(\omega)|^2 d\omega , 
\end{eqnarray}
where $\mathcal{L}_1(\omega) = - j \alpha \omega / (\alpha - j\omega)^2$ is the Fourier transform of $w(t)L_1(t)$. From~(\ref{Eq:Mod-noise-average-variance}) and using $k_w = 1/2$, we obtain the average variance of the modelling error for ERLS algorithms: 
\begin{eqnarray} \label{Eq:mod_average_variance_ERLS} 
\sigma_\text{mod}^2 = \frac{P \sigma_s^2}{4\pi} \int_{-\infty}^{\infty} G_h(\omega) \frac{\alpha^2 \omega^2}{(\alpha^2 + \omega^2)^2} d\omega .   
\end{eqnarray}
Using the approximation $(1+x)^{-2} \approx 1 - 2x$~\cite{dwight1947tables} in the last integral and equation~(\ref{Eq:NMSD_epsilon_n_ERLS}) for conversion of the modelling noise into the corresponding NMSD component, for random processes with finite $\lambda_2$ we obtain 
\begin{eqnarray} \label{Eq:mod_ERLS} 
\epsilon_\text{m}^\text{(ERLS)} \approx  \frac{\lambda_2}{2 \alpha^2}    \frac{(1 - \lambda) L }{(1 + \lambda) } . 
\end{eqnarray}
This formula and its variants for the uniform and Jakes' PSDs are shown in Table~\ref{Table:ERLS_NMSD_components}.

Note that for the AR random process with the PSD $G_h(\omega)$ in~(\ref{Eq:AR_Gh}) the spectral moment $\lambda_2$ does not exist and therefore the approximation in~(\ref{Eq:mod_ERLS}) is not valid. In this case, the direct integration in~(\ref{Eq:mod_average_variance_ERLS}) results~in           
\begin{eqnarray} 
\epsilon_\text{m}^\text{(ERLS)} & \approx & \frac{\mu \alpha}{4(\mu + \alpha)^2} \frac{(1 - \lambda) L }{(1 + \lambda) } \label{Eq:mod_error_AR_ERLS_1}  \\
&=& \frac{ \ln(\lambda) \ln(\beta)}{4[\ln(\lambda) + \ln(\beta)]^2} \frac{(1 - \lambda) L }{(1 + \lambda) } .   \label{Eq:mod_error_AR_ERLS_2} 
\end{eqnarray}

\begin{table*}
\caption{Approximation and modelling NMSD components of ERLS and dERLS algorithms}
\label{Table:ERLS_NMSD_components}
\centering
\begin{tabular}{|c|ccc|}
\hline
NMSD component                                                          & \multicolumn{1}{c|}{\begin{tabular}[c]{@{}c@{}}Classical ERLS\\ ($t_0 = 0$)\end{tabular}} & \multicolumn{1}{c|}{\begin{tabular}[c]{@{}c@{}}dERLS\\ (arbitrary $t_0$)\end{tabular}}                                                                      & \begin{tabular}[c]{@{}c@{}}Optimal dERLS\\ ($t_0 = \alpha^{-1}$)\end{tabular} \\ \hline
\begin{tabular}[c]{@{}c@{}}Approximation\\ (general form)\end{tabular}  & \multicolumn{1}{c|}{$\frac{\lambda_2}{\alpha^2} + \frac{\lambda_4}{\alpha^4}$}            & \multicolumn{1}{c|}{$\frac{\lambda_2}{\alpha^2}(1 - \alpha t_0)^2 + \frac{\lambda_4}{\alpha^4} \left(1 - \alpha t_0 + \frac{\alpha^2 t_0^2}{2} \right)^2 $} & $\frac{\lambda_4}{4 \alpha^4}$                                                \\ \hline
\begin{tabular}[c]{@{}c@{}}Approximation\\ (Uniform PSD)\end{tabular}   & \multicolumn{1}{c|}{$\frac{\xi^2}{3}  + \frac{\xi^4}{5}  $}         & \multicolumn{1}{c|}{$\frac{\xi^2}{3}(1 - \alpha t_0)^2 + \frac{\xi^4}{5} \left(1 - \alpha t_0 + \frac{\alpha^2 t_0^2}{2} \right)^2 $}                                                                                                                                       &   $\frac{\xi^4}{20}$                                                                              \\ \hline
\begin{tabular}[c]{@{}c@{}}Approximation \\ (Jakes' PSD)\end{tabular} & \multicolumn{1}{c|}{$\frac{\xi^2}{2}  + \frac{3  \xi^4}{8} $}         & \multicolumn{1}{c|}{$\frac{\xi^2}{2}(1 - \alpha t_0)^2 + \frac{3 \xi^4}{8} \left(1 - \alpha t_0 + \frac{\alpha^2 t_0^2}{2} \right)^2 $}                                                                                                                                       &    $\frac{3 \xi^4}{32}$                                                                               \\ \hline
\begin{tabular}[c]{@{}c@{}}Approximation \\ (AR process)\end{tabular} & \multicolumn{1}{c|}{$\frac{\mu}{\alpha + \mu} = \frac{\ln(\beta)}{\ln(\beta \lambda)}$}         & \multicolumn{1}{c|}{$-$}                                                                                                                                       &   $-$                                                                            \\ \hline
\begin{tabular}[c]{@{}c@{}}Modelling \\ (general form)\end{tabular}     & \multicolumn{3}{c|}{$\frac{\lambda_2}{2\alpha^2} \cdot \frac{(1-\lambda)L}{1+\lambda}$                                                                                                                                                                                                                                          } \\ \hline
\begin{tabular}[c]{@{}c@{}}Modelling\\ (Uniform PSD)\end{tabular}       & \multicolumn{3}{c|}{$ \frac{\xi^2}{6}  \cdot \frac{(1-\lambda)L}{1+\lambda} $ }                                                                                                                                                                                                                                                                                      \\ \hline
\begin{tabular}[c]{@{}c@{}}Modelling\\ (Jakes' PSD)\end{tabular}      & \multicolumn{3}{c|}{$ \frac{\xi^2}{4} \cdot \frac{(1-\lambda)L}{1+\lambda}$}                                                                                                                                                                                                                                                                                      \\ \hline
\begin{tabular}[c]{@{}c@{}}Modelling\\ (AR process)\end{tabular}      & \multicolumn{3}{c|}{$\frac{\mu \alpha}{4(\mu + \alpha)^2} \cdot \frac{(1-\lambda)L}{1+\lambda} = \frac{\ln(\lambda) \ln(\beta)}{4[\ln(\lambda) + \ln(\beta)]^2} \cdot \frac{(1-\lambda)L}{1+\lambda}$}                                                                                                                                                                                                                                                                                      \\ \hline
\end{tabular}
\end{table*}

\subsection{Approximation NMSD component for ERLS algorithms \label{Subsec:Approx-Error-ERLS}}

\subsubsection{Approximation NMSD component for the classical ERLS algorithm} 

For the channel variations represented as the Taylor series~(\ref{Eq:h_Taylor}), from~(\ref{Eq:h_Laguerre_expansion_coefficients}) we obtain
\begin{eqnarray} \label{Eq:ERLS_h_hat}
\hat{h}_\ell(0) =  \alpha \sum_{p = 0}^{\infty} \frac{h_\ell^{(p)}(0)}{p!} \int_{-\infty}^{0} t^p e^{\alpha t} dt.  
\end{eqnarray}
The integral in~(\ref{Eq:ERLS_h_hat}) is given by~\cite{dwight1947tables}
\begin{eqnarray} \label{Eq:ERLS_integral}
\int_{-\infty}^{0} t^p e^{\alpha t} dt = (-1)^p \frac{p!}{\alpha^{p+1}}  . 
\end{eqnarray}
Therefore, the estimate in~(\ref{Eq:ERLS_h_hat}) can be written as:
\begin{eqnarray} \label{Eq:ERLS_h_hat_final}
\hat{h}_\ell(0) =  \sum_{p = 0}^{\infty} \frac{(-1)^p h_\ell^{(p)}(0)}{\alpha^p} .  
\end{eqnarray}
The term $p = 0$ in~(\ref{Eq:ERLS_h_hat_final}) is $h_\ell(0)$, therefore the estimation error $e_{\text{ERLS},\ell} = \hat{h}_\ell(0) - h_\ell(0)$ can be written as:
\begin{eqnarray} \label{Eq:ERLS_error}
e_{\text{ERLS},\ell} =  \sum_{p = 1}^{\infty} \frac{(-1)^p h_\ell^{(p)}(0)}{\alpha^p} .  
\end{eqnarray}
      
To find the variance of the error $e_{\text{ERLS},\ell}$, $E \{ e^2_{\text{ERLS},\ell}\}$, we keep the first two terms in~(\ref{Eq:ERLS_error}) and take into account that $h^{(1)}(0)$ and $h^{(2)}(0)$ are uncorrelated; as a result, we obtain 
\begin{eqnarray} \label{Eq:ERLS_MSD_single_tap}
E \{ e^2_{\text{ERLS},\ell}\} \approx P_\ell \left( \frac{\lambda_2}{\alpha^2} + \frac{\lambda_4}{\alpha^4} \right).
\end{eqnarray}
Finally, from (\ref{Eq:NMSD}), we arrive at the approximation NMSD component for the classical ERLS algorithm: 
\begin{eqnarray} \label{Eq:ERLS_NMSD}
\epsilon_{\text{a
}}^{\text{(ERLS)}} \approx  \frac{\lambda_2}{\alpha^2} + \frac{\lambda_4}{\alpha^4} .
\end{eqnarray}
This formula and its variants for the uniform and Jakes' PSDs are shown in Table~\ref{Table:ERLS_NMSD_components}, where we take into account that $\alpha = -\ln{(\lambda)}/T_s$ and denote $\xi = 2 \pi f_m / \ln{\lambda}$. Note that the first term in~(\ref{Eq:ERLS_NMSD}) can also be obtained from an integral MSD expression in~\cite{lin1995optimal}, where the first term of $\epsilon_\text{a}^\text{(ERLS)}$ for Jakes' PSD has also been obtained in the same form as shown in Table~\ref{Table:ERLS_NMSD_components}.


We now consider the approximation error for the AR process. In this case, we cannot use the representation (\ref{Eq:h_Taylor}) since this process is not differentiable. Instead, we need to directly consider 
\begin{eqnarray} \label{Eq:ERLS_AR_sigma_h}
E\{e_{\text{ERLS},\ell}^2\} = E\left\{ \left[ h_\ell(0) - \alpha \int_{-\infty}^{0} e^{\alpha t} h_\ell(t) dt \right]^2 \right\}. 
\end{eqnarray}
This can be represented as $E\{e_{\text{ERLS},\ell}^2\} = P_\ell - J_1 + J_2$, where $J_1$ and $J_2$ are integrals defined below. First we compute 
\begin{eqnarray} \label{Eq:ERLS_AR_J1}
J_1 = 2 \alpha \int_{-\infty}^{0} e^{\alpha \tau} \rho_h(\tau) d\tau = \frac{2 \alpha P_\ell}{\alpha + \mu},  
\end{eqnarray}
where $\rho_h(\tau) = P_\ell e^{-\mu |\tau|}$ is the correlation function of the process $h_\ell(t)$. 
The second integral $J_2$ is given by
\begin{eqnarray} \label{Eq:ERLS_AR_J2}
J_2 = \alpha^2 \int_{-\infty}^{0} \int_{-\infty}^{0} e^{\alpha (t + \tau)} \rho_h(t - \tau) dt d\tau = \frac{\alpha P_\ell}{\alpha + \mu}.   
\end{eqnarray}
By combining results in~(\ref{Eq:ERLS_AR_J1}) and~(\ref{Eq:ERLS_AR_J2}), for the AR process, we obtain
\begin{eqnarray} \label{Eq:ERLS_AR_sigma_h_final}
E\{e_{\text{ERLS},\ell}^2\} = \frac{\mu P_\ell}{\alpha + \mu}. 
\end{eqnarray}
Finally, using (\ref{Eq:NMSD}) and (\ref{Eq:ERLS_AR_sigma_h_final}), we arrive at the approximation NMSD component
\begin{eqnarray} \label{Eq:ERLS_AR_approx_error}
\epsilon_\text{a}^\text{(ERLS)} = \frac{\mu}{\alpha + \mu}. 
\end{eqnarray}
Taking into account that $\lambda = e^{-\alpha T_s}$ and $\beta = e^{-\mu T_s}$, we obtain this component in terms of the parameters $\lambda$ (forgetting factor of the ERLS algorithm) and $\beta$ (the memory factor of the channel time variations) as
\begin{eqnarray} \label{Eq:ERLS_AR_approx_error_lambda_beta}
\epsilon_\text{a}^\text{(ERLS)} = \frac{\ln(\beta)}{\ln(\beta \lambda)}. 
\end{eqnarray}

\subsubsection{Approximation NMSD component for the dERLS algorithm}

We assume that, without additive noise, the dERLS algorithm for a time-varying tap $h(t)$, at a time $t = 0$, computes the estimate
\begin{eqnarray} \label{Eq:dERLS_h_hat_integral}
\hat{h}(0) =  \alpha \int_{-\infty}^{t_0} e^{\alpha (t-t_0)} h(t) dt . 
\end{eqnarray}
In this case, for the channel variations represented by the Taylor series~(\ref{Eq:h_Taylor}), we obtain
\begin{eqnarray} \label{Eq:dERLS_h_hat}
\hat{h}_\ell(0) =  \sum_{p = 0}^{\infty} \frac{h_\ell^{(p)}(0)}{p!} \left[ \alpha \int_{-\infty}^{t_0} t^p e^{\alpha (t - t_0)} dt \right].  
\end{eqnarray}
The integral in~(\ref{Eq:dERLS_h_hat}) can be computed as: 
\begin{eqnarray} \label{Eq:dERLS_integral}
\alpha \int_{-\infty}^{t_0} t^p e^{\alpha t} dt = p! \sum_{k = 0}^{p} (-1)^{p-k} \frac{t_0^k}{k! \alpha^{p-k}} . 
\end{eqnarray}
Therefore, the estimate in~(\ref{Eq:dERLS_h_hat}) can be written as:
\begin{eqnarray} \label{Eq:dERLS_h_hat_final}
\hat{h}_\ell(0) =  \sum_{p = 0}^{\infty} \frac{(-1)^p h_\ell^{(p)}(0)}{\alpha^p} \sum_{k = 0}^{p} (-1)^{k} \frac{ \alpha^{k} t_0^k}{k!} .  
\end{eqnarray}
The term $p = 0$ in~(\ref{Eq:dERLS_h_hat_final}) is $h_\ell(0)$, therefore the estimation error $e_{\text{dERLS},\ell} = \hat{h}_\ell(0) - h_\ell(0)$ can be written as:
\begin{eqnarray} \label{Eq:dERLS_error}
e_{\text{dERLS},\ell} = \sum_{p = 1}^{\infty} \frac{(-1)^p h_\ell^{(p)}(0)}{\alpha^p} \sum_{k = 0}^{p} (-1)^{k} \frac{ \alpha^{k} t_0^k}{k!}   .  
\end{eqnarray}

The first two terms in~(\ref{Eq:dERLS_error}) provide an approximation
\begin{eqnarray} \label{Eq:dERLS_error_two_terms}
e_{\text{dERLS},\ell} \approx &-& \frac{h_\ell^{(1)}(0)}{\alpha} (1 - \alpha t_0) \nonumber \\
&+&  \frac{h_\ell^{(2)}(0)}{\alpha^2} \left(1 - \alpha t_0 + \frac{\alpha^2 t_0^2}{2} \right)      .  
\end{eqnarray}
It can be seen that the error due to the first derivative (linear channel variation) can be completely eliminated by setting the delay $t_0 = \alpha^{-1}$, and, in this case, the error is given by  
\begin{eqnarray} \label{Eq:dERLS_error_one_term_optimal}
e_{\text{dERLS},\ell} \approx \frac{h_\ell^{(2)}(0)}{2 \alpha^2}  .  
\end{eqnarray}
The error variance $ E \{ e^2_{\text{dERLS},\ell} \}$ for the optimal delay $t_0 = \alpha^{-1}$ is then given by
\begin{eqnarray} \label{Eq:dERLS_MSD_single_tap-optimal-delay}
E \{ e^2_{\text{dERLS},\ell} \} \approx P_\ell \frac{\lambda_4}{4 \alpha^4} .
\end{eqnarray}
Finally, the approximation NMSD component for the dERLS algorithm with the optimal delay $t_0 = \alpha^{-1}$ is then given by 
\begin{eqnarray} \label{Eq:dERLS_MSD_optimal_delay}
\epsilon_{a}^{\text{dERLS}} \approx \frac{\lambda_4}{4 \alpha^4} .
\end{eqnarray}
For other values of $t_0$, we obtain
\begin{eqnarray} \label{Eq:dERLS_MSD_single_tap}
\varepsilon_{\text{dERLS}}  \approx  \frac{\lambda_2}{\alpha^2} (1 - \alpha t_0)^2 
+  \frac{\lambda_4}{\alpha^4} \left(1 - \alpha t_0 + \frac{\alpha^2 t_0^2}{2} \right)^2 .
\end{eqnarray}
In the case $t_0 = 0$, from~(\ref{Eq:dERLS_MSD_single_tap}) we obtain the approximation NMSD component~(\ref{Eq:ERLS_NMSD}) for the ERLS algorithm. 

Table~\ref{Table:ERLS_NMSD_components} summarises results of this section.

\section{Numerical results} \label{Sec:Numerical-results}

In our simulation examples, we focus on scenarios, in which the additive noise is not present; the known formulas~(\ref{Eq:NMSD_epsilon_n_SRLS}) and~(\ref{Eq:NMSD_epsilon_n_ERLS}) can predict the noise NMSD component.

\subsection{SRLS, dSRLS and SRLS-L algorithms} 

We first consider numerical examples for SRLS algorithms. 

\begin{figure}[h]
    \centering
    \begin{subfigure}[b]{0.5\textwidth}
        \centering
        \includegraphics[height=1.7in]{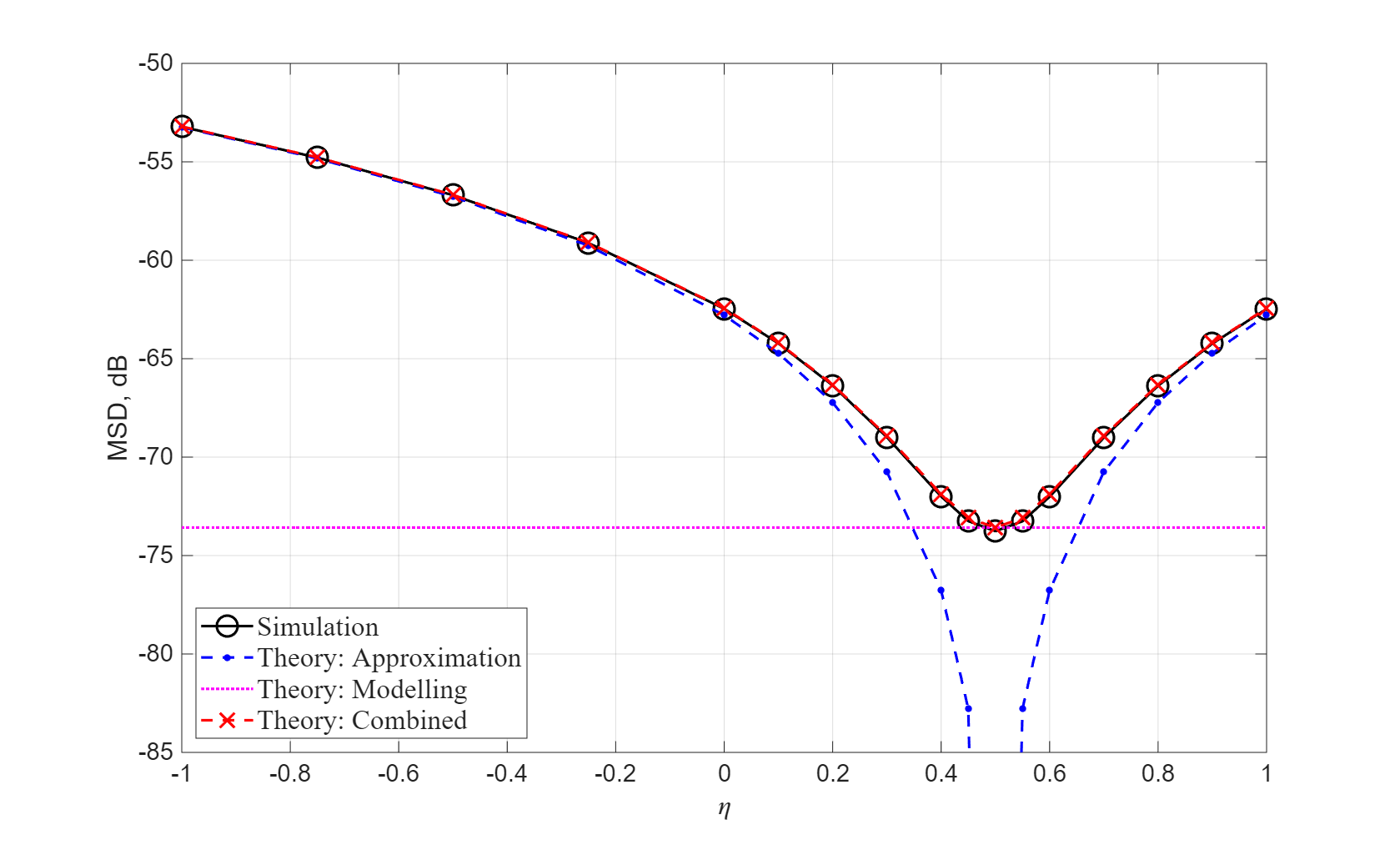}
        \caption{Uniform PSD, uniform PDP, $f_m = 10^{-5}$}
    \end{subfigure}
    ~
    \begin{subfigure}[b]{0.5\textwidth}
        \centering
        \includegraphics[height=1.7in]{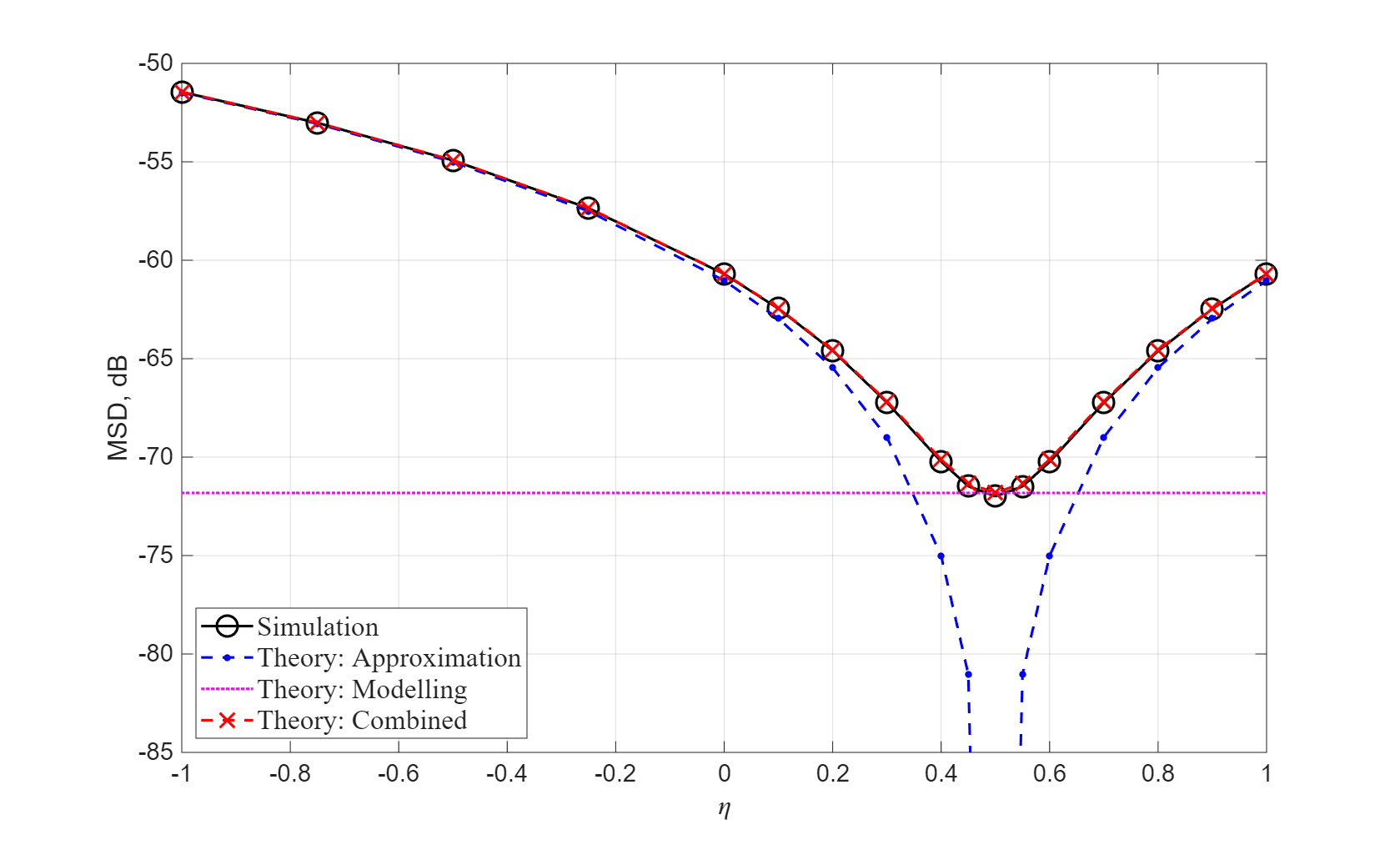}
        \caption{Jakes' PSD, uniform PDP, $f_m = 10^{-5}$}
    \end{subfigure}
    ~
    \begin{subfigure}[b]{0.5\textwidth}
        \centering
        \includegraphics[height=1.7in]{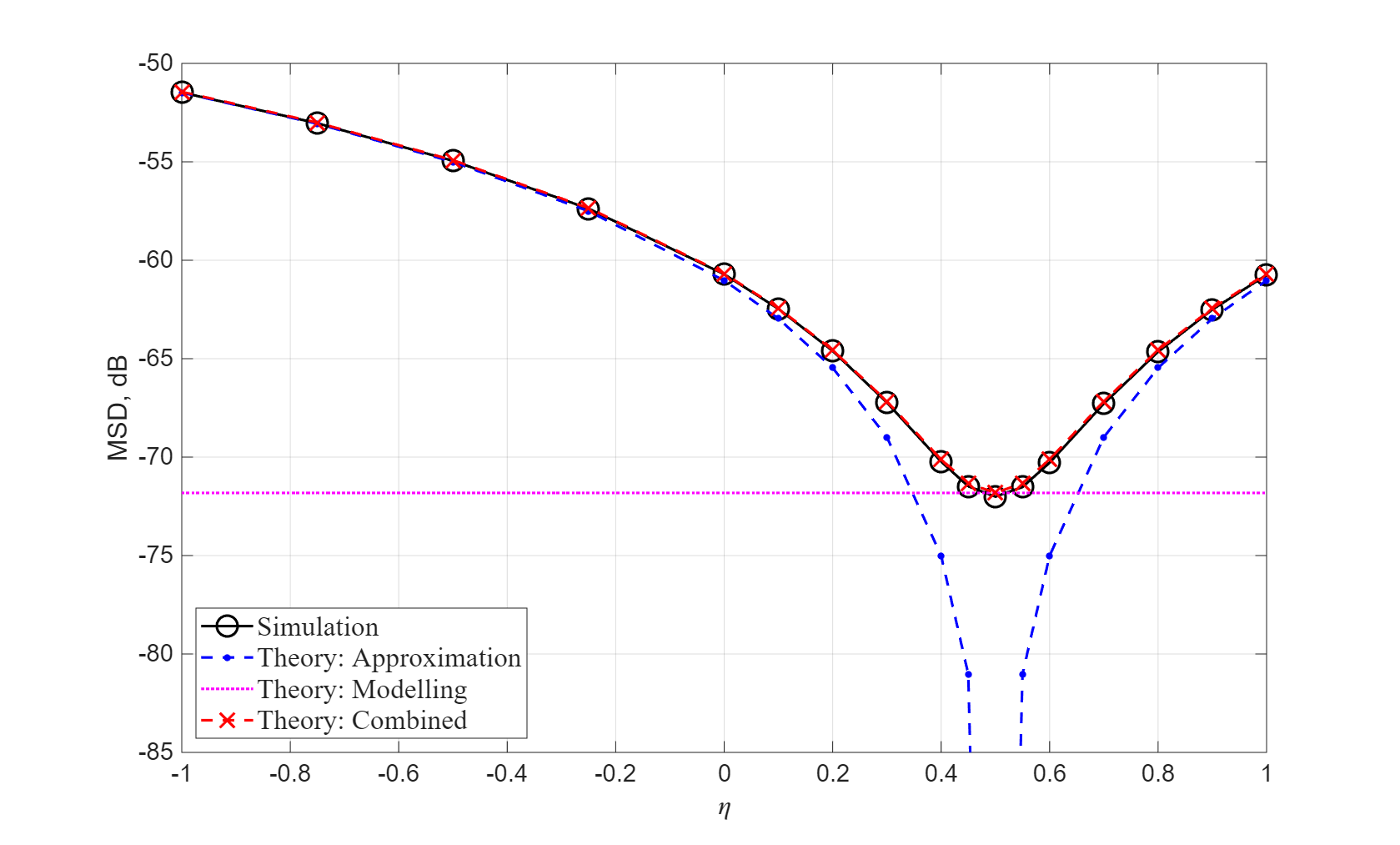}
        \caption{Jakes' PSD, triangle PDP, $f_m = 10^{-5}$}
    \end{subfigure}
    ~
    \begin{subfigure}[b]{0.5\textwidth}
        \centering
        \includegraphics[height=1.7in]{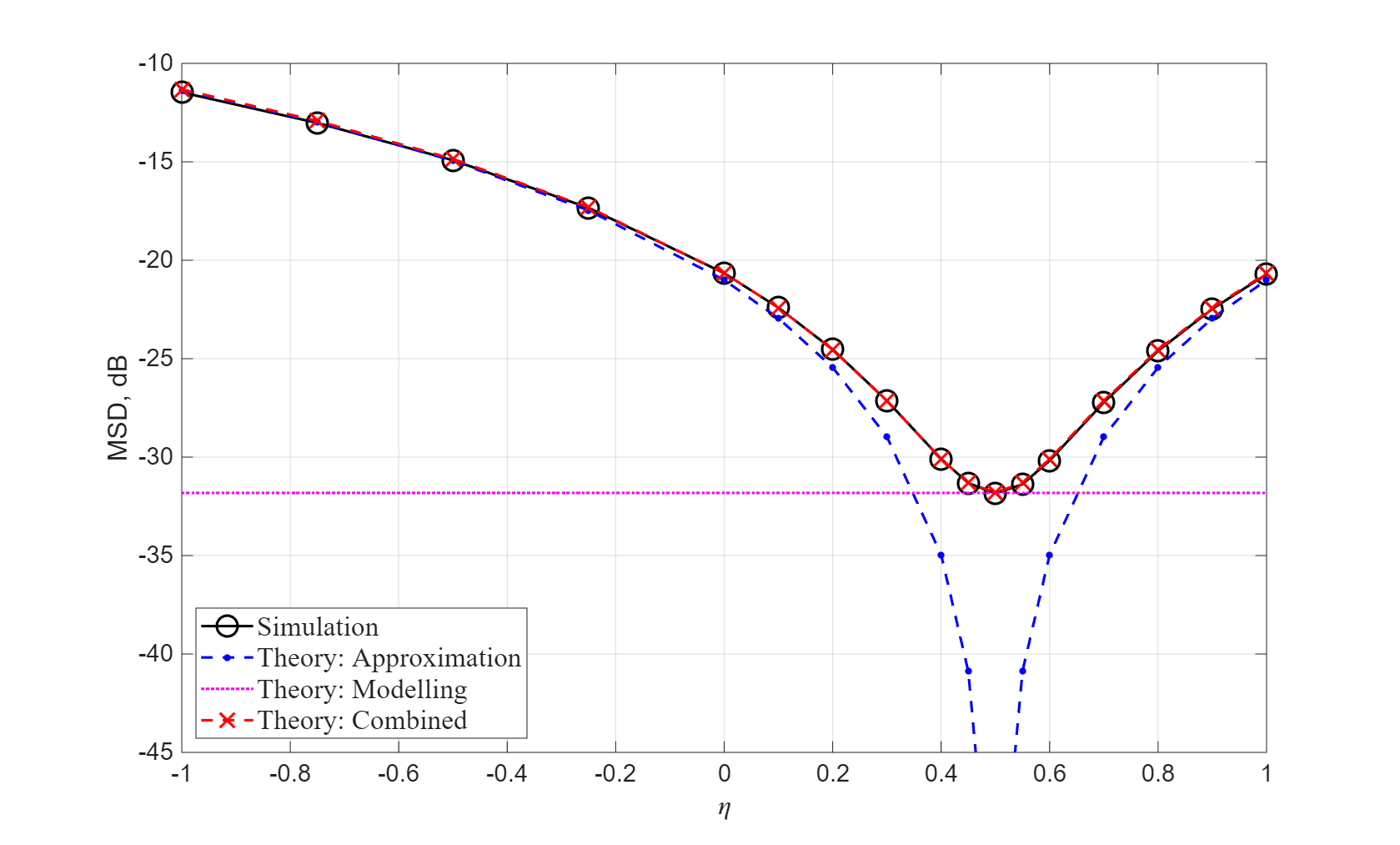}
        \caption{Jakes' PSD, triangle PDP, $f_m = 10^{-3}$}
    \end{subfigure}
    \caption{\label{Fig:NMSD-dSRLS_Triangle_PDP} NMSD performance of the dSRLS algorithm against the normalized delay $\eta$ in the time-varying multipath channel: $L = 10$, $M = 40$.}
\end{figure}
Fig.~\ref{Fig:NMSD-dSRLS_Triangle_PDP} presents the dSRLS performance against the delay $\eta = t_0 / T$ for the case $L = 10$, specifically $\epsilon_\text{m}$, $\epsilon_\text{a}$, and $\epsilon_\text{am} = \epsilon_\text{a} + \epsilon_\text{m}$ computed according to formulas in Table~\ref{Table:SRLS_NMSD_components} and corresponding simulation results. Fig.~\ref{Fig:NMSD-dSRLS_Triangle_PDP}(a) and Fig.~\ref{Fig:NMSD-dSRLS_Triangle_PDP}(b) show these dependencies for the uniform and Jakes' PSDs. It is seen that in the vicinity of the optimal delay, the NMSD is mostly defined by the modelling component, while away from the optimal delay, the approximation component provides an accurate prediction; the combined NMSD obtained from the theoretical results in Table~\ref{Table:SRLS_NMSD_components} provides accurate prediction of the numerical results for any $t_0$; the maximum discrepancy between the theoretical and numerical results is about 0.2~dB, which is observed at the optimal delay $\eta = 0.5$; at other $\eta$, it is smaller, and for the classical SRLS algorithm ($\eta = 0$), it is less than 0.03~dB. It is also seen that the NMSD for the uniform PSD is lower than that for Jakes' PSD; according to Table~\ref{Table:SRLS_NMSD_components}, the ratio between them is approximately 1.5, i.e. 1.8~dB, consistent with what is observed in Fig.~\ref{Fig:NMSD-dSRLS_Triangle_PDP}(a) and Fig.~\ref{Fig:NMSD-dSRLS_Triangle_PDP}(b).    
Fig.~\ref{Fig:NMSD-dSRLS_Triangle_PDP}(c) shows results for the triangle PDP. Comparison with Fig.~\ref{Fig:NMSD-dSRLS_Triangle_PDP}(b) confirms that NMSD does not depend on PDP. Fig.~\ref{Fig:NMSD-dSRLS_Triangle_PDP}(d) shows results for significantly faster channel variations ($f_m = 10^{-3}$ against $f_m = 10^{-5}$ in the other subfigures). It is seen that the theoretical prediction is still accurate with a maximum discrepancy of less than 0.15~dB at $\eta = -1$; for other delays the discrepancy is significantly lower.

\begin{figure}
\begin{center}  
\includegraphics[width=0.5\textwidth]{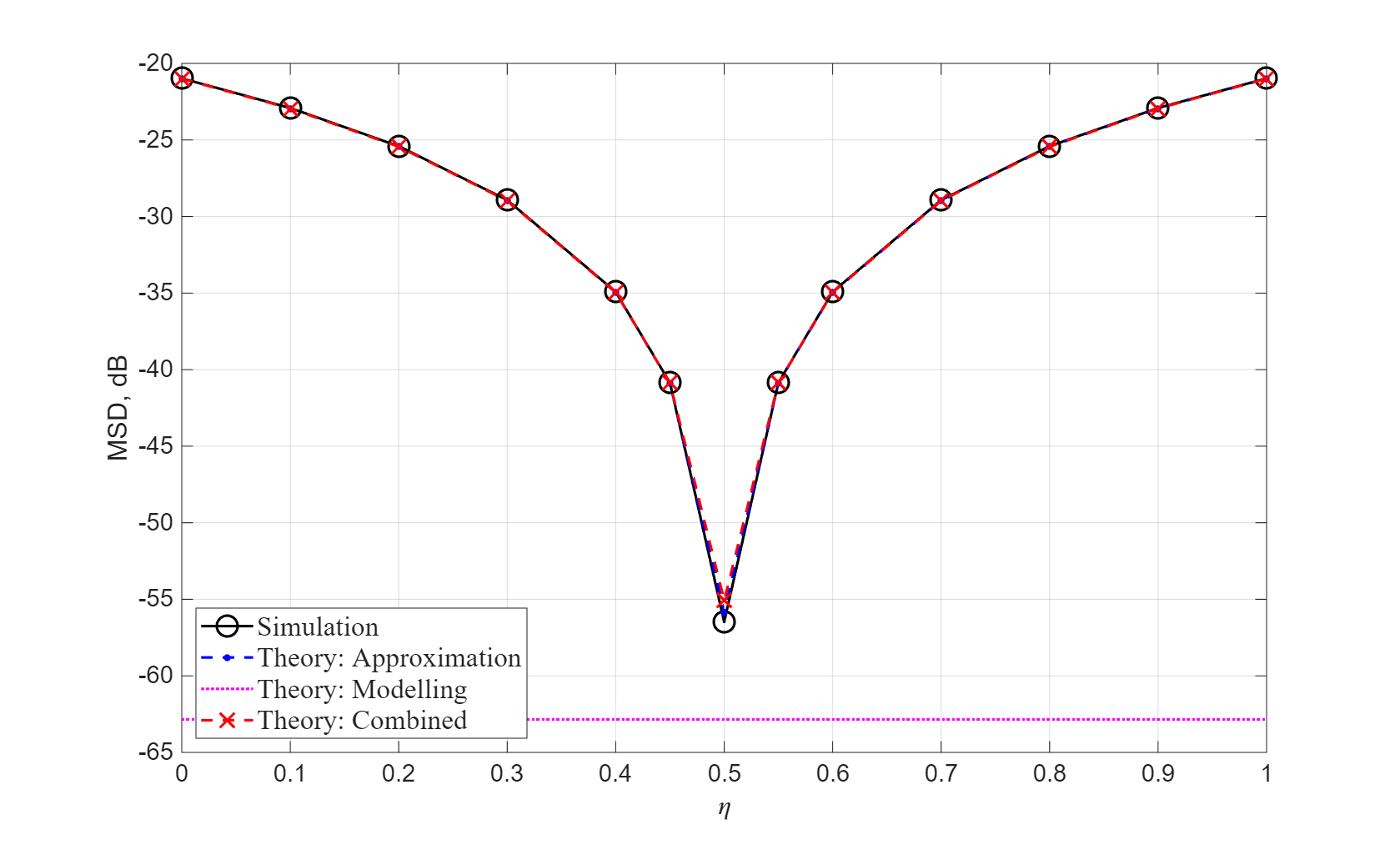}
\caption{\label{Fig:SRLS_L_TPDP_JPSD_fm3_v19} MSD performance of the SRLS-L algorithm against the delay $\eta$ in the multipath channel, $L = 10$,  $M = 40$, Jakes' PSD, triangle PDP, $f_m = 10^{-3}$. } 
\end{center}
\end{figure}
Fig.~\ref{Fig:SRLS_L_TPDP_JPSD_fm3_v19} shows results for the SRLS-L algorithm in channels with $f_m = 10^{-3}$. In this case, the modelling error is almost negligible, so the tracking performance is dominated by the approximation error, and we can see a good match between the theoretical and numerical results. The highest discrepancy 1.4~dB is observed at the optimal delay $\eta = 0.5$, at the other values of $\eta$ shown in Fig.~\ref{Fig:SRLS_L_TPDP_JPSD_fm3_v19}, it is smaller than 0.04~dB. Comparison with results in Fig.~\ref{Fig:NMSD-dSRLS_Triangle_PDP}(d) shows that, at the optimal delay ($\eta = 0.5$), the tracking performance of the SRLS-L algorithm is significantly better compared to the dSRLS algorithm, the improvement is about 24~dB.

\begin{figure}
\begin{center}  
\includegraphics[width=0.5\textwidth]{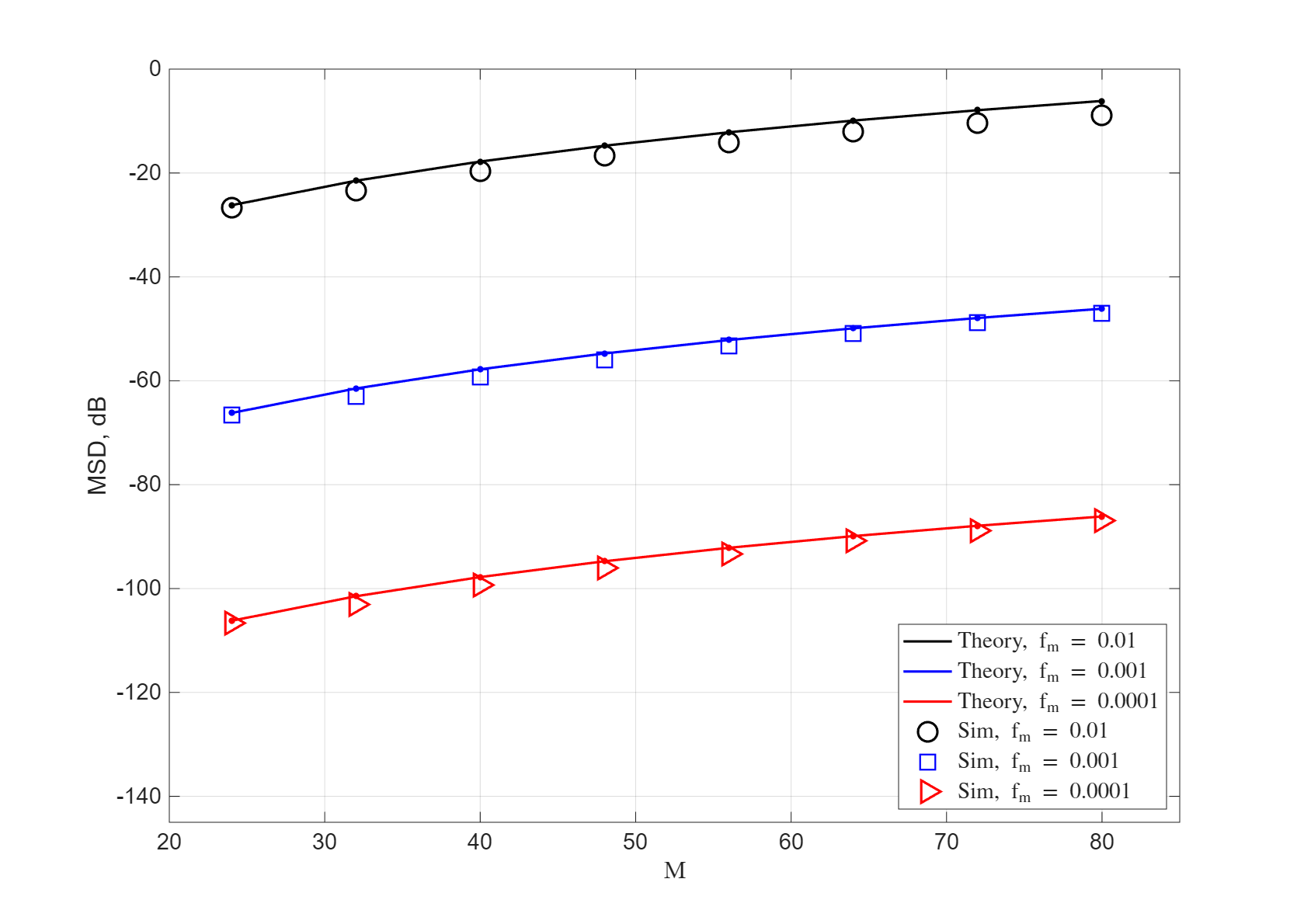}
\caption{\label{Fig:MSD-SRLS-L-L10-Theory-Sim-1} MSD performance of the SRLS-L algorithm against the window length $M$ in the multipath channel ($L = 10$) with Jakes' PSD and uniform PDP; $\eta = 0.5$. } 
\end{center}
\end{figure}
Fig.~\ref{Fig:MSD-SRLS-L-L10-Theory-Sim-1} shows the SRLS-L performance against the window length $M$ for the optimal delay ($i_0 = M/2$). We can again observe good match between the theoretical prediction and numerical results. Higher discrepancies are observed in the fast-varying channel ($f_m = 0.01$) at longer $M$; note that, in such cases, the parameter $\gamma$ takes values in the interval $\gamma \in [0.24, 0.8]$ with $\gamma = 0.8$ for $M = 80$, i.e., the relationship $\gamma \ll 1$ is not satisfied. For such values of $\gamma$, the linear approximation of channel variations over the observation window is not accurate and more Taylor series terms (e.g., parabolic) are required for a better channel representation.

\subsection{ERLS and dERLS algorithms} 

\begin{figure}[t!]
    \centering
    \begin{subfigure}[b]{0.5\textwidth}
        \centering
        \includegraphics[height=2in]{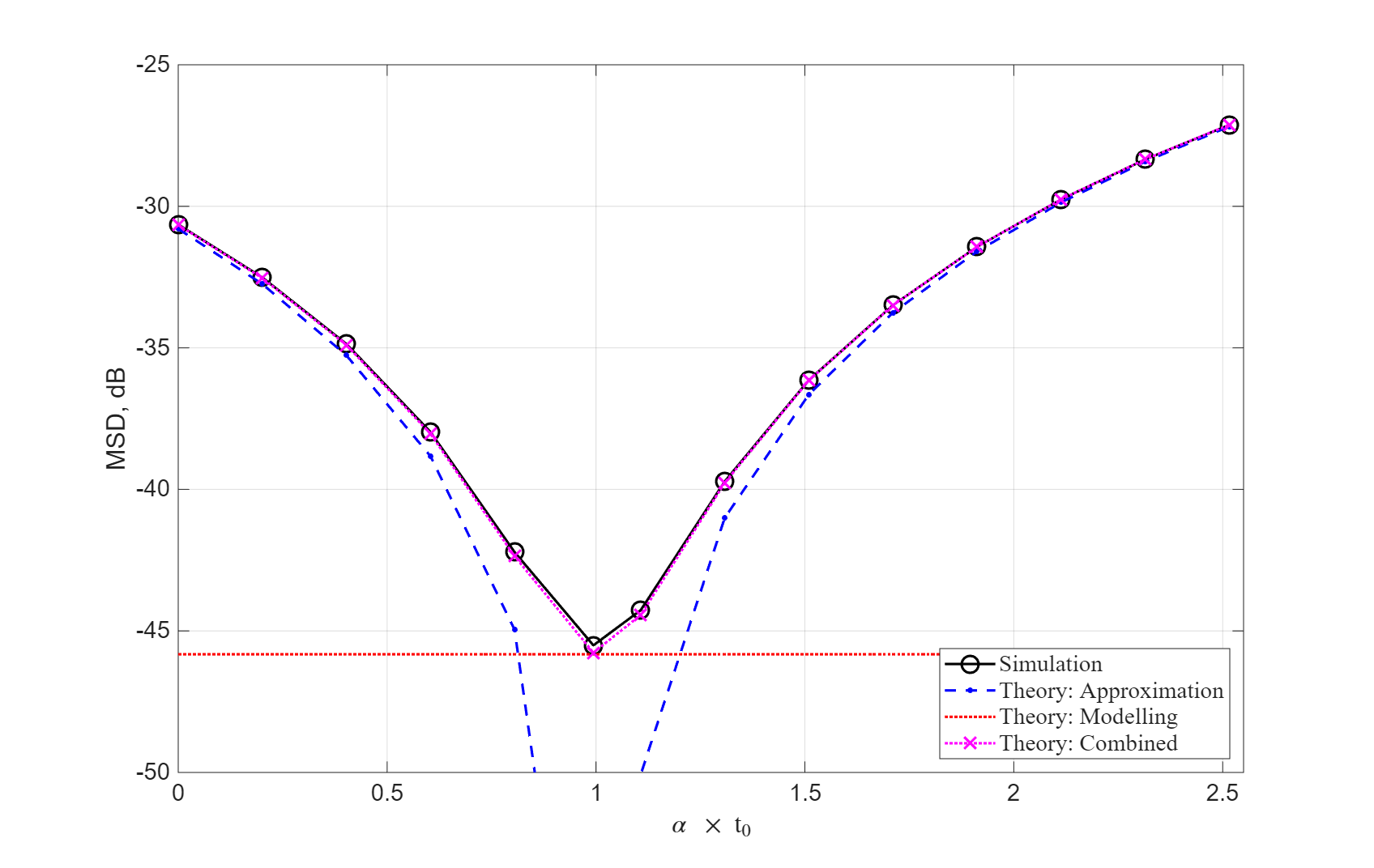}
        \caption{$\lambda = 1 - 1/(8L) = 0.9875 $}
    \end{subfigure}
    ~ 
    \begin{subfigure}[b]{0.5\textwidth}
        \centering
        \includegraphics[height=2in]{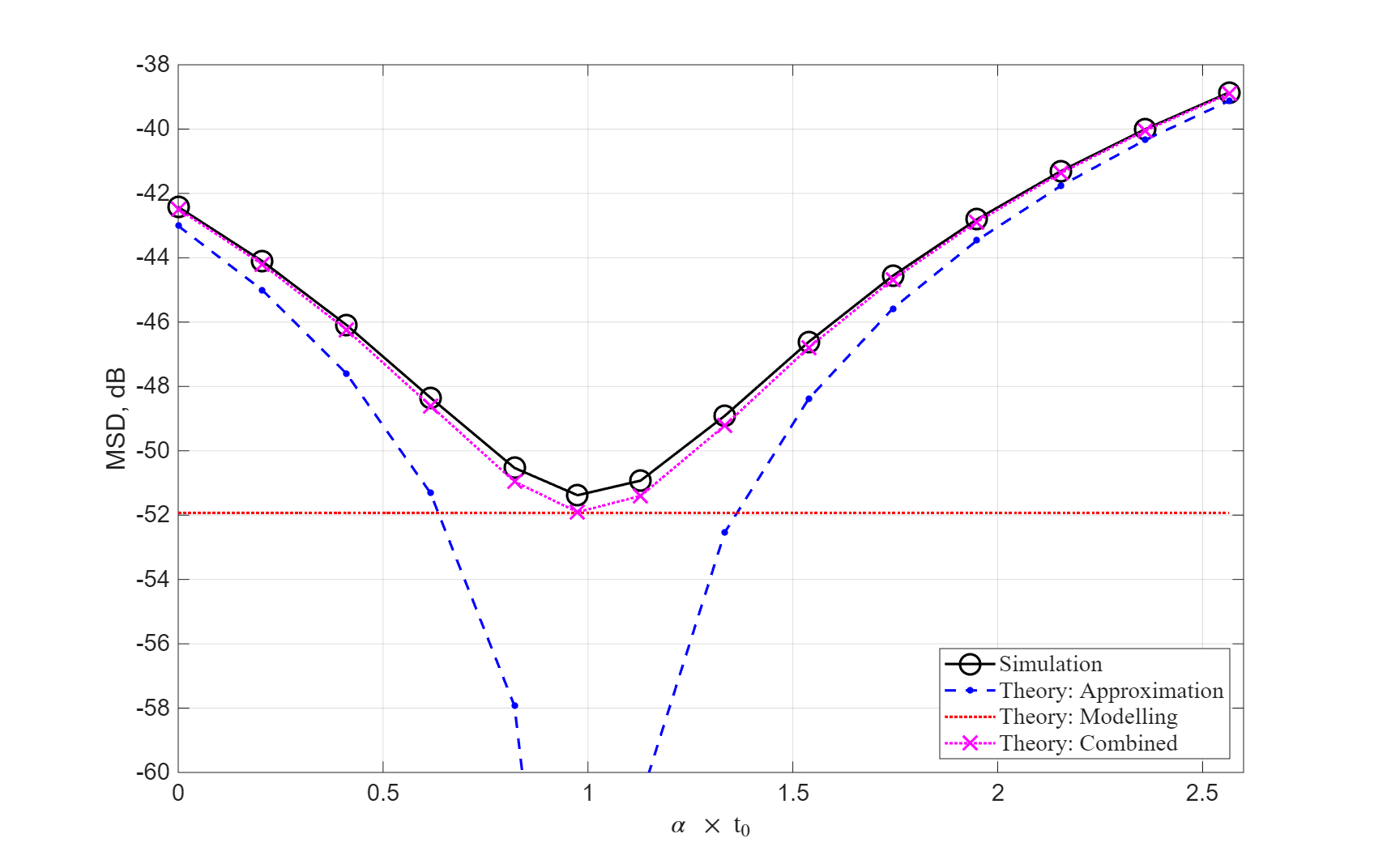}
        \caption{$\lambda = 1 - 1/(2L) = 0.95$}
    \end{subfigure}
    \caption{\label{Fig:NMSD-dERLS_UPSD_UPDP_against_lambda} NMSD performance of the dERLS algorithm against the delay $t_0$ in the time-varying multipath channels ($L = 10$) with the uniform PDP and uniform PSD with $f_m = 10^{-4}$.}
\end{figure}
Fig.~\ref{Fig:NMSD-dERLS_UPSD_UPDP_against_lambda} shows the MSD performance of the dERLS algorithm against the delay for two values of the forgetting factor~$\lambda$. It can be seen that, for $\lambda$ close to 1 in Fig.~\ref{Fig:NMSD-dERLS_UPSD_UPDP_against_lambda}(a), the theoretical and simulation results are very close: a (maximum) discrepancy of 0.3~dB is observed at $t_0$ near the optimal delay ($\alpha t_0 \approx 1$, $i_0 = 79$). However, for the classical ERLS algorithm ($t_0 = 0$), the discrepancy is as small as 0.02~dB. For smaller $\lambda$ ($\lambda = 0.95$), i.e., a narrower observation window, the (maximum) discrepancy in the vicinity of the optimal delay increases to 0.5~dB; this can be explained by the fact that for smaller $\lambda$, the integral approximation in~(\ref{Eq:dERLS_h_hat_integral}) of dERLS estimates is less accurate. However, with such a small $\lambda$, $\lambda = 0.95$, for the classical ERLS algorithm ($t_0 = 0$), the discrepancy, while slightly increased (compared to the case $\lambda = 0.9875$), remains only 0.06~dB.

\begin{figure}
\begin{center}  
\includegraphics[width=0.5\textwidth]{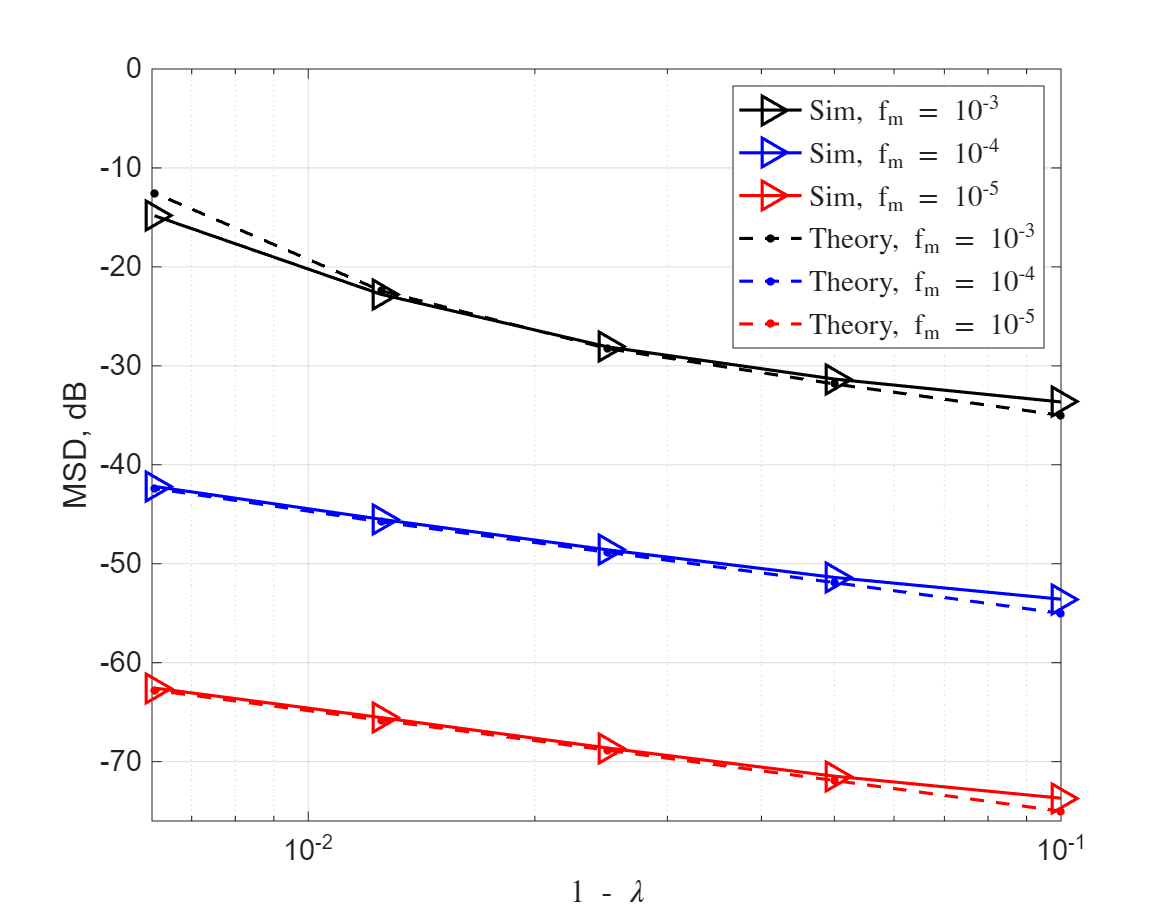}
\caption{\label{Fig:dERLS_UPDP_UPSD_opt_t0_against_lambda} The MSD performance of the dERLS algorithm against the forgetting factor $\lambda$ for the channel with the uniform PDP and uniform PSD: theoretical and numerical results in the multipath channel ($L = 10$). The delay is set close to the optimal delay $t_0 = \alpha^{-1}$; more specifically, $(\alpha T_s)^{-1} = - 1 / \ln{\lambda}$ is rounded to the nearest integer number. For the forgetting factors $\lambda = 0.9, 0.95, 0.975, 0.9875, 0.9938$ used in this simulation example, the delays are $i_0 = 9, 19, 39, 79, 159$, respectively.} 
\end{center}
\end{figure}
Since the maximum discrepancy between the theoretical and numerical results is observed in the vicinity of the optimal delay, in Fig.~\ref{Fig:dERLS_UPDP_UPSD_opt_t0_against_lambda} we investigate the accuracy of the performance prediction at a delay close to the optimal delay at different $\lambda$ and speed of channel variations. It can be seen that the prediction improves as $\lambda$ approaches 1 and speed of the channel variations reduces. For the smallest $\lambda$, $\lambda = 0.9$, the discrepancy is about 1.4~dB for all the channel variation speeds, and with increase in $\lambda$ it becomes significantly smaller. An exception is for the highest speed of channel variations ($f_m = 10^{-3}$) and highest $\lambda = 0.9938$, when the discrepancy is about 2.3~dB; in this case, the efficient time window of the adaptive algorithm is about~$320$ samples, which is comparable with the channel time coherence, about $1/f_m = 1000$ samples. The prediction accuracy can be improved if more Taylor series terms are included in the derivation of the modelling and approximation NMSD components.           

\begin{figure}
\begin{center}  
\includegraphics[width=0.5\textwidth]{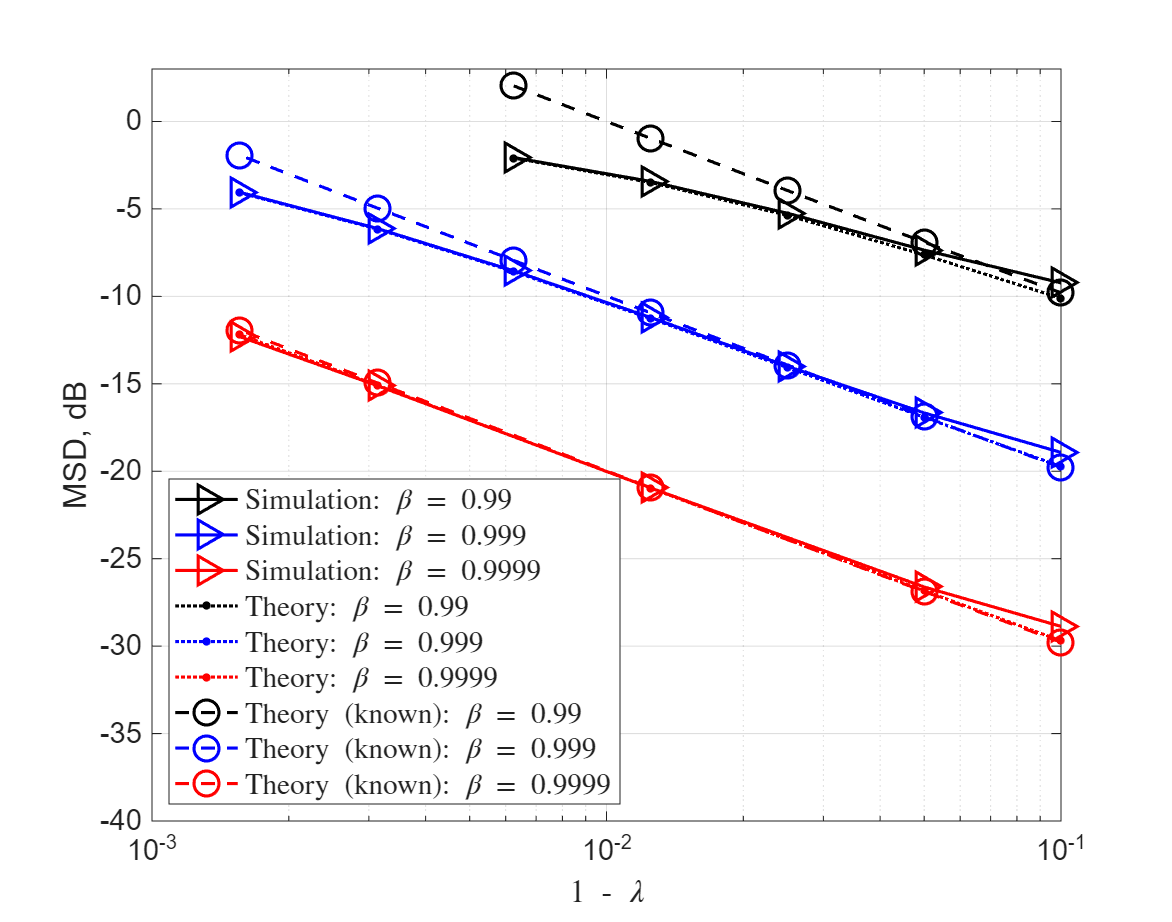}
\caption{\label{Fig:ERLS_UPDP_AR_PSD_v8d} MSD performance of the ERLS algorithm ($t_0 = 0$) against the forgetting factor $\lambda$ for the channel with the uniform PDP and AR PSD: theoretical and numerical results in the multipath channel ($L = 10$). }
\end{center}
\end{figure}
Fig.~\ref{Fig:ERLS_UPDP_AR_PSD_v8d} shows similar results for the AR PSD channel time variations described by the parameter $\beta$. It can be seen that the formulas derived in this paper more accurately predict the ERLS tracking performance in comparison with the prediction $\epsilon_\text{am} \approx (1 - \beta^2)/(1-\lambda^2)$, which follows from results in~\cite{eleftheriou1986tracking}.

\section{Conclusions} \label{Sec:Conclusions}

In this paper, we have proposed a novel approach for analysing the tracking performance of RLS adaptive algorithms in WSSUS channels.  The normalised MSD is expressed as a sum of three components: noise, modelling and approximation components. Previous works addressing the tracking performance of classical RLS algorithms, such as the ERLS and SRLS algorithms, did not account for the modelling component caused by the channel variations. This was justified since the modelling component for the classical RLS algorithms is relatively low compared with the approximation component. However, for RLS algorithms with improved tracking performance, such as the dSRLS and dERLS algorithms, in the vicinity of the optimal delay, the modelling component dominates the tracking performance, therefore known approaches are no longer valid.       

The proposed approach has been used to derive simple formulas for prediction of the MSD performance of the classical RLS algorithms against the algorithm parameters and channel characteristics, such as spectral moments of the PSD of channel variations. They have been specified for uniform, Jakes' and AR PSDs. Some of the expressions of the approximation component are similar to those previously derived in the literature. Our MSD formulas provide similar or improved match with numerical simulation results, when compared to known results. The proposed approach has also enabled us to derive simple expressions for the MSD of the dSRLS, dERLS and SRLS-L adaptive algorithms; these expressions have been shown to provide a good match with numerical results in a variety of simulation scenarios.  

When analysing the MSD performance, we tried to avoid operations with matrices, since they significantly complicate the analysis. Therefore, a number of idealizations were introduced as follows. 
\begin{enumerate}
        \item The discrete-time calculations, involved in the operation of the adaptive algorithms, have been replaced with continuous-time operations. The replacement of the sums by integrals introduces errors that we ignore in this paper. An analysis of these errors is required to better justify such a step.  
        \item The tap decoupling has been assumed in the derivations. Although, this is widely true for time-invariant systems, this assumption requires a justification for the time-variant case, which is the case of interest in this paper.   
        \item The fact that derivatives of stationary Gaussian random processes are uncorrelated has been exploited in the derivations. This allows at most two terms of the Taylor series to be kept in the analysis. Although two terms have been enough for scenarios considered in the paper, for faster channel variations, more terms can be useful. 
        \item The non-stationary modelling noise has been replaced with a stationary noise with a variance equal to the average variance of the non-stationary noise, plugged into equations~(\ref{Eq:NMSD_epsilon_n_SRLS}) and~(\ref{Eq:NMSD_epsilon_n_ERLS}) derived for a stationary noise. Such a treatment of the modelling noise requires a justification or a replacement with a more rigorous step.            
\end{enumerate}
Also, for the analysis of the AR process of the first order, the derived general formulas with the spectral moments cannot be used, since the required spectral moments do not exist. Still, the direct integration in our analysis has resulted in more accurate (compared to known) expressions for calculating the MSD performance. 

However, these idealizations have resulted in simple and accurate formulas for prediction of tracking performance of multiple RLS adaptive algorithms used for estimation of time-varying impulse responses of linear systems with the parameter variations described as WSSUS processes. 
Future work will focus on validating these idealizations/assumptions.

\bibliography{RLS_tracking}

@article{zakharov2004polynomial,
  title={{Polynomial spline-approximation of Clarke's model}},
  author={Zakharov, Yuriy V and Tozer, Tim C and Adlard, Jonathan F},
  journal={IEEE Transactions on Signal Processing},
  volume={52},
  number={5},
  pages={1198--1208},
  year={2004},
  publisher={IEEE}
}

@article{karami2007tracking,
  title={Tracking performance of least squares {MIMO} channel estimation algorithm},
  author={Karami, Ebrahim},
  journal={IEEE Transactions on Communications},
  volume={55},
  number={11},
  pages={2201--2209},
  year={2007},
  publisher={IEEE}
}

@inproceedings{bershad1990performance,
  title={Performance comparison of {RLS} and {LMS} algorithms for tracking a first order {M}arkov communications channel},
  author={Bershad, NJ and McLaughlin, S and Cowan, CFN},
  booktitle={IEEE International Symposium on Circuits and Systems},
  pages={266--270},
  year={1990}
}

@article{lin1995optimal,
  title={Optimal tracking of time-varying channels: {A} frequency domain approach for known and new algorithms},
  author={Lin, Jingdong and Proakis, John G. and Ling, Fuyun and Lev-Ari, Hanoch},
  journal={IEEE Journal on Selected Areas in Communications},
  volume={13},
  number={1},
  pages={141--154},
  year={1995}
}

@inproceedings{guo1992tracking,
  title={Tracking performance analysis of the forgetting factor {RLS} algorithm},
  author={Guo, Lei and Ljung, Lennart and Priouret, Pierre},
  booktitle={Proceedings of the 31st IEEE Conference on Decision and Control},
  pages={688--693},
  year={1992}
}

@article{zheng2003simulation,
  title={Simulation models with correct statistical properties for {R}ayleigh fading channels},
  author={Zheng, Yahong Rosa and Xiao, Chengshan},
  journal={IEEE Transactions on Communications},
  volume={51},
  number={6},
  pages={920--928},
  year={2003},
  publisher={IEEE}
}

@article{xiao2002second,
  title={Second-order statistical properties of the {WSS J}akes' fading channel simulator},
  author={Xiao, Chengshan and Zheng, Yahong R and Beaulieu, Norman C},
  journal={IEEE Transactions on Communications},
  volume={50},
  number={6},
  pages={888--891},
  year={2002},
  publisher={IEEE}
}

@article{zheng2002improved,
  title={Improved models for the generation of multiple uncorrelated {R}ayleigh fading waveforms},
  author={Zheng, Yahong R and Xiao, Chengshan},
  journal={IEEE Communications Letters},
  volume={6},
  number={6},
  pages={256--258},
  year={2002},
  publisher={IEEE}
}

@article{brillinger1974fourier,
  title={Fourier analysis of stationary processes},
  author={Brillinger, David R},
  journal={Proceedings of the IEEE},
  volume={62},
  number={12},
  pages={1628--1643},
  year={1974},
  publisher={IEEE}
}

@inproceedings{shen2021performance,
  title={Performance of adaptive filtering based on {L}egendre polynomials},
  author={Shen, Lu and Zakharov, Yuriy and Shi, Long and Henson, Benjamin},
  booktitle={IEEE Statistical Signal Processing Workshop (SSP)},
  pages={6--10},
  year={2021}
}

@article{kay1993statistical,
  title={Statistical signal processing: {E}stimation theory},
  author={Kay, Steven M},
  journal={Prentice Hall},
  year={1993}
}

@inproceedings{filimon1993lms,
  title={{LMS and RLS tracking analysis for WSSUS channels}},
  author={Filimon, Voicu and Kozek, Werner and Kreuzer, Werner and Kubin, Gernot},
  booktitle={IEEE International Conference on Acoustics, Speech, and Signal Processing},
  volume={3},
  pages={348--351},
  year={1993}
}

@article{haykin2002adaptive,
  title={Adaptive tracking of linear time-variant systems by extended {RLS} algorithms},
  author={Haykin, Simon and Sayed, Ali H and Zeidler, James R and Yee, Paul and Wei, Paul C},
  journal={IEEE Transactions on Signal Processing},
  volume={45},
  number={5},
  pages={1118--1128},
  year={1997},
  publisher={IEEE}
}

@article{silva2008improving,
  title={Improving the tracking capability of adaptive filters via convex combination},
  author={Silva, Magno TM and Nascimento, Vitor H},
  journal={IEEE Transactions on Signal Processing},
  volume={56},
  number={7},
  pages={3137--3149},
  year={2008},
  publisher={IEEE}
}

@article{eweda1994comparison,
  title={Comparison of {RLS}, {LMS}, and sign algorithms for tracking randomly time-varying channels},
  author={Eweda, Eweda},
  journal={IEEE Transactions on Signal Processing},
  volume={42},
  number={11},
  pages={2937--2944},
  year={1994},
  publisher={IEEE}
}

@article{claser2021tracking,
  title={On the tracking performance of adaptive filters and their combinations},
  author={Claser, Raffaello and Nascimento, Vitor H},
  journal={IEEE Transactions on Signal Processing},
  volume={69},
  pages={3104--3116},
  year={2021},
  publisher={IEEE}
}

@article{eleftheriou1986tracking,
  title={Tracking properties and steady-state performance of {RLS} adaptive filter algorithms},
  author={Eleftheriou, Evangelos and Falconer, D},
  journal={IEEE Transactions on Acoustics, Speech, and Signal Processing},
  volume={34},
  number={5},
  pages={1097--1110},
  year={1986},
  publisher={IEEE}
}

@article{dwight1947tables,
  title={Tables of integrals and other mathematical data},
  author={Dwight, Herbert Bristol},
  journal={New York: The MacMillan Company},
  year={1947}
}

@article{feng2008statistical,
  title={Statistical analysis of mobile radio reception: An extension of Clarke's model},
  author={Feng, Tao and Field, Timothy R},
  journal={IEEE Transactions on Communications},
  volume={56},
  number={12},
  pages={2007--2012},
  year={2008},
  publisher={IEEE}
}

@book{sayed2003fundamentals,
  title={Fundamentals of adaptive filtering},
  author={Sayed, Ali H},
  year={2003},
  publisher={John Wiley \& Sons}
}

@book{leadbetter2012extremes,
  title={Extremes and related properties of random sequences and processes},
  author={Leadbetter, Malcolm R and Lindgren, Georg and Rootz{\'e}n, Holger},
  year={2012},
  publisher={Springer Science \& Business Media}
}

@book{cramer2013stationary,
  title={Stationary and related stochastic processes: Sample function properties and their applications},
  author={Cram{\'e}r, Harald and Leadbetter, M Ross},
  year={2013},
  publisher={Courier Corporation}
}

@article{niedzwiecki2019generalized,
  title={Generalized {S}avitzky--{G}olay filters for identification of nonstationary systems},
  author={Nied{\'z}wiecki, Maciej and Cio{\l}ek, Marcin},
  journal={Automatica},
  volume={108},
  pages={108477},
  year={2019},
  publisher={Elsevier}
}

@article{shen2022bem,
  title={{BEM} adaptive filtering for {SI} cancellation in full-duplex underwater acoustic systems},
  author={Shen, Lu and Zakharov, Yuriy and Shi, Long and Henson, Benjamin},
  journal={Signal Processing},
  volume={191},
  pages={108366},
  year={2022},
  publisher={Elsevier}
}

@article{shen2019digital,
title={Digital self-interference cancellation for underwater acoustic systems},
  author={Shen, Lu and Henson, Benjamin and Zakharov, Yuriy and Mitchell, Paul},
  journal={IEEE Transactions on Circuits and Systems II: Express Briefs},
  volume={67},
  number={1},
  pages={192--196},
  year={2020},
  publisher={IEEE}
}

@article{shen2020adaptive,
  title={Adaptive filtering for full-duplex {UWA} systems with time-varying self-interference channel},
  author={Shen, Lu and Zakharov, Yuriy and Henson, Benjamin and Morozs, Nils and Mitchell, Paul D},
  journal={IEEE Access},
  volume={8},
  pages={187590--187604},
  year={2020},
  publisher={IEEE}
}

@article{zhang2018soft,
  title={Soft-decision-driven sparse channel estimation and turbo equalization for {MIMO} underwater acoustic communications},
  author={Zhang, Youwen and Zakharov, Yuriy V and Li, Jianghui},
  journal={IEEE Access},
  volume={6},
  pages={4955--4973},
  year={2018},
  publisher={IEEE}
}

@inproceedings{zakharov2016sliding,
  title={Sliding window adaptive filter with diagonal loading for estimation of sparse {UWA} channels},
  author={Zakharov, Yuriy V and Li, Jianghui},
  booktitle={IEEE OCEANS, Shanghai},
  pages={1--5},
  year = {2016}
}

@article{yeo2000improved,
  title={Improved {RLS} algorithm for time-variant underwater acoustic communications},
  author={Yeo, HK and Sharif, BS and Hinton, OR and Adams, AE},
  journal={Electronics Letters},
  volume={36},
  number={2},
  pages={191--192},
  year={2000},
  publisher={IET}
}

@book{gradshteyn2014table,
  title={Table of integrals, series, and products},
  author={Gradshteyn, I.S. and Ryzhik, I.M.},
  isbn={9781483265643},
  year={2014},
  publisher={Elsevier Science}
}

@article{percival1993simulating,
  title={Simulating {G}aussian random processes with specified spectra},
  author={Percival, Donald B},
  journal={Computing Science and Statistics},
  pages={534--534},
  year={1993}
}
\bibliographystyle{IEEEtran}

\end{document}